\documentclass[12pt]{article}
\usepackage[utf8]{inputenc}
\usepackage[svgnames,dvipsnames]{xcolor}
\usepackage[a4paper,left=1.0cm,right=1.0cm,top=2.0cm,bottom=2.0cm]{geometry}
\usepackage{empheq}
\usepackage[bottom, multiple]{footmisc}
\usepackage{booktabs, multirow, nicefrac, float, xspace}
\usepackage{graphicx,subcaption}
\usepackage{tikz,placeins}
\usetikzlibrary{decorations.markings}
\usetikzlibrary{decorations.pathmorphing}
\usepackage{framed}
\usepackage{csquotes}
\definecolor{shadecolor}{rgb}{0.90,0.90,0.90}
\usepackage{cite}
\usepackage{hyperref}
\usepackage{subcaption}
\usepackage{pifont}
\usepackage{mathtools}
\usepackage{setspace}
\usepackage{amsmath, amssymb, amsthm, float, graphicx,amsfonts,braket}
\numberwithin{equation}{section}
\usepackage{amsthm}

\theoremstyle{definition}

\usepackage{float}
\allowdisplaybreaks

\hypersetup{colorlinks=true, linkcolor=Maroon, citecolor=FireBrick,urlcolor=Green,linktocpage}
\definecolor{amber(sae/ece)}{rgb}{1.0, 0.49, 0.0}
\definecolor{deepsaffron}{rgb}{1.0, 0.6, 0.2}

\usepackage{multirow}

\usepackage{amstext} 
\usepackage{array}   
\newcolumntype{L}{>{$}l<{$}}
\newcolumntype{C}{>{$}c<{$}}

\def\beq{\begin{eqnarray}}\def\eeq{\end{eqnarray}}
\def\be{\begin{equation}}\def\ee{\end{equation}}

\def\s{\sigma}

\def\l{\lambda}

\usepackage{bm}
\usepackage{bm}

\def\kernel{{\mathcal H}^{(\l)}(\s,s,t)}
\def\disptau{\tau^{(\lambda)}(\sigma,s,t)}

\begin{document}

\title{\bf A stringy dispersion relation\\ for field theory}
\author{Faizan Bhat$^{a}$\footnote{faizanbhat@iisc.ac.in}, Arnab Priya Saha$^{a}$\footnote{arnabsaha@iisc.ac.in}~~and Aninda Sinha$^{a,b}$\footnote{asinha@iisc.ac.in}\\
\it ${^a}$Centre for High Energy Physics,
\it Indian Institute of Science,\\ \it C.V. Raman Avenue, Bangalore 560012, India.\\ \it ${^b}$Department of Physics and Astronomy, University of Calgary,\\ \it Alberta T2N 1N4, Canada.}

\maketitle

\abstract{We derive a local, crossing symmetric dispersion relation (CSDR) for 2-2 scattering amplitudes with a parametric ambiguity motivated by string theory. Various limits of the parameter lead to the fixed-$t$, fixed-$s$, and other known CSDRs. We also present formulae for higher-subtracted cases. Several examples are discussed for illustration. In particular, for the Veneziano and the Virasoro-Shapiro amplitudes, we derive parametric series representations which manifest poles in all channels and converge everywhere. We then discuss applications of our formalism for bootstrapping weakly-coupled gravitational EFTs. We demonstrate that even in the presence of the graviton pole, one can derive bounds on the Wilson coefficients while working in the forward limit, with the parameter acting as the IR regulator instead. Finally, we derive series representations for multi-variable, totally symmetric generalisations of the Veneziano and Virasoro-Shapiro amplitudes that manifest poles in all the variables. This is a first step towards dispersion relations for $n$-particle scattering amplitudes.}

\newpage
\tableofcontents
\onehalfspacing

\section{Introduction}
\begin{figure}[H]
    \centering
    \includegraphics[width= 0.8\linewidth]{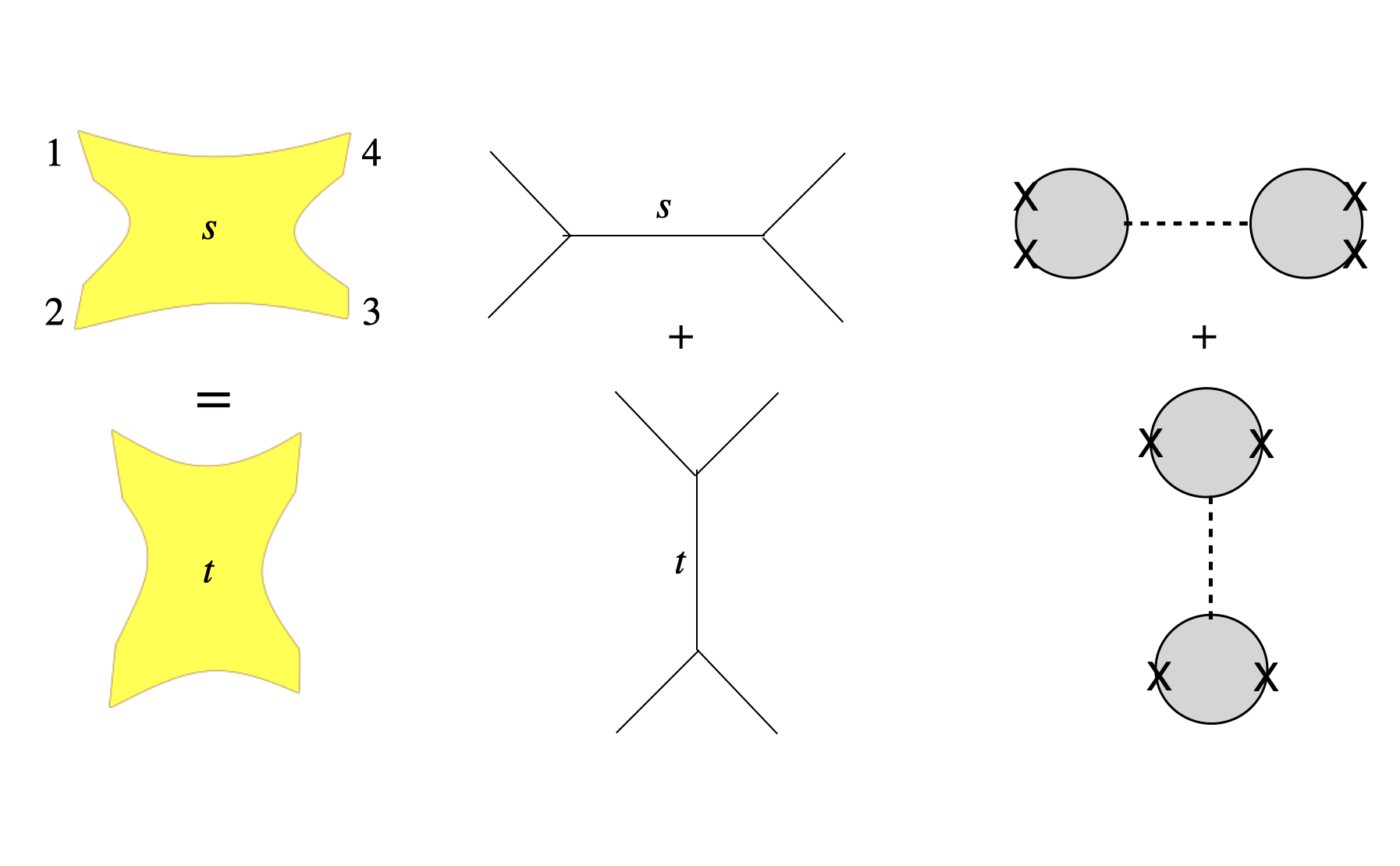}
    \caption{The open string worldsheet picture on the left, Feynman diagrams in the middle and the product of disc representations of the channels (black crosses depict vertex operators).}
    \label{fig:sdisp}
\end{figure}

The tree-level 2-2 scattering amplitudes in string theory \cite{polchinski} admit multiple series representations. For example, consider the Veneziano amplitude. The most well-known representations are expansions in terms of the $s$-channel or the $t$-channel poles. 
\begin{equation}
\label{VenAmp}
\begin{split}
\frac{\Gamma(-s)\Gamma(-t)}{\Gamma(- s - t)}
&= \sum_{n=0}^{\infty} \frac{-1}{n!} \,
\frac{(t + n )(t + n - 1)\cdots (t + 1)}{s - n},
\qquad \mathrm{Re}(t) < 0 \\
&= \sum_{n=0}^{\infty} \frac{-1}{n!} \,
\frac{(s + n)(s + n - 1)\cdots (s + 1)}{t - n},
\qquad \mathrm{Re}(s) < 0
\end{split}
\end{equation}
The equivalence of these two representations is known as the "dual resonance model" hypothesis \cite{mandelstam, birth}. The hypothesis is seemingly at odds with the perturbative quantum field theory Feynman diagram picture, where the amplitude is represented as a sum over poles in both channels. In the older literature, a representation with poles in both channels (referred to as the interference model) was considered incorrect as it would lead to ``double counting". However, string field theory \cite{polchinski, sen} suggests that there should exist a representation of string amplitudes similar to the Feynman diagram expansion, involving a sum over all channels and possibly extra contact diagram terms. Furthermore, explicit calculations in \cite{sen} indicate that such a representation may also have a parametric ambiguity, which in the plumbing fixture construction reflects the freedom to choose coordinates on the worldsheet around the vertex operator insertions. In quantum field theories, this is related to the usual field redefinition ambiguity. Another way to motivate the parametric ambiguity is from the string worldsheet picture depicted in Fig. \ref{fig:sdisp}. On the left, the worldsheet picture suggests that there should be some ``elasticity parameter" that stretches the worldsheet into the form of the $s$-channel or $t$-channel, hinting at the existence of a representation using which one can ``deform" the $t$-channel form to the $s$-channel form. On the right, the disc diagram picture arising from the plumbing fixture construction suggests that the same representation should give a Feynman diagram-like expansion. Therefore, an important task is to explicitly derive such representations for 2-2 string amplitudes. 

Representations of the kind given in \eqref{VenAmp} can be derived using dispersion relations for 2-2 scattering amplitudes \cite{hagedorn, martin1}. The widely used fixed-$t$ dispersion relation leads to the first equation in \eqref{VenAmp}, while the fixed-$s$ dispersion relation gives the second one. In recent times, dispersion relations have been extensively used to study constraints on EFTs (for a partial list of references see \cite{snow}). Since the fixed-$t/s$ dispersion relations only make poles in one channel manifest, the representations for scattering amplitudes typically have a restricted domain of convergence. For example, in \eqref{VenAmp}, we have a restricted domain in $t$ because the $t$-poles of the amplitude are not manifest in the dispersive representation. These appear only after analytic continuation (see \cite{mizera} for a recent discussion).

In \cite{SS}, a new parametric representation of the Veneziano amplitude was presented that exhibits poles in all channels and converges everywhere else for all $s, t$.
\begin{equation}
\label{VenAmpPCSDR}
\frac{\Gamma(-s_1)\Gamma(-s_2)}{\Gamma(- s_1 - s_2)}
= \sum_{n=0}^{\infty} \frac{-1}{n!}
\left(
\frac{1}{s_1 - n} + \frac{1}{s_2 - n} + \frac{1}{\lambda + n}
\right)
\left(
(1 - \lambda) + \frac{(s_1 + \lambda)(s_2 + \lambda)}{n+\lambda}
\right)_{n} ,
\end{equation}
where $(x)_n$ is the Pochammer symbol. The representation is independent of the parameter $\lambda$ and converges when $\text{Re}(\lambda) > 0$. We see that it matches the string field theory expectations outlined above\footnote{Disclaimer: It is not obvious how the parameter is exactly related to that in string field theory.}. This motivates the search for such parametric representations for tree-level string amplitudes in general.

To arrive at representations of the kind in \eqref{VenAmpPCSDR}, we first need a dispersion relation that is manifestly crossing symmetric. In the 1970s, a new way of writing crossing symmetric dispersion relations (CSDR) was introduced in  \cite{old1, old2}. However, this approach did not garner much attention until recently, when in \cite{SZ, GSZ}, it was revived and then applied to a large class of problems in QFTs and CFTs \cite{appl1,appl2,appl3,appl4,appl5,appl6,appl7, appl8, appl9,appl10,appl11,appl12,appl13}. The penalty one pays is that the CSDR has some spurious terms that lead to non-local features. In \cite{SZ, GSZ}, these non-local terms were subtracted by hand, and a manifestly local basis for expanding non-perturbative scattering amplitudes was constructed, analogously to the Feynman diagrams in flat space and Witten diagrams in AdS. In \cite{song}, Song implemented the same subtraction procedure for non-local terms in a more elegant way and derived a remarkably simple CSDR that naturally led to this local basis. Finally, in \cite{appl4}, starting from the Song's local CSDR, a new CSDR was presented that is local and also has a one-parameter ambiguity, in accordance with string field theory expectations. We refer to this as parametric or stringy CSDR. Using the parametric CSDR, one can derive representations like \eqref{VenAmpPCSDR} \footnote{In \cite{rosengren}, mathematician Hjalmar Rosengren found an alternative proof of such representations of Veneziano-like amplitudes}. However, the complete derivation of the stringy CSDR, starting from the original derivation of the CSDR in \cite{old1} (see \cite{SZ} for a brief review), is quite involved.  This is because the derivation of the original CSDR itself requires a complicated parametrisation of the Mandelstam variables in terms of new variables, then writing the dispersion relation in those new variables, and finally, reverting to the Mandelstam variables. This is followed by Song's subtraction procedure and then the tricks used in \cite{appl4} to finally arrive at the parametric CSDR. This lengthy procedure is not very illuminating. In this paper, we address this shortcoming by providing a straightforward, self-contained derivation of the parametric CSDR, similar in spirit to how the fixed-$t$ dispersion relation is derived using Cauchy's residue theorem. This provides a general non-perturbative representation of 2-2 scattering amplitudes with a parametric ambiguity. When applied to string amplitudes in particular, it allows us to derive representations of the kind in \eqref{VenAmpPCSDR} and its generalizations to the three-channel case.

The parametric CSDR also finds applications in the S-matrix bootstrap program. In fact, one of the main motivations behind pursuing these new representations is to develop a more efficient basis for the numerical bootstrap. In \cite{appl4}, we presented evidence that the parametric dispersive representation enables us to impose bootstrap constraints on a wider region of the kinematical variable space. In \cite{appl5,appl6,appl7}, several interesting applications of the CSDR have been identified, showcasing some of its advantages over conventional approaches. In this work, we explain that the parametric CSDR proves especially advantageous for bootstrapping weakly-coupled gravitational EFTs. In \cite{Swamp}, the low-energy Wilson coefficients that parametrise these EFTs were shown to be bounded in flat space. Due to the presence of the graviton pole, the usual way of deriving bounds on the Wilson coefficients that uses the fixed-$t$ dispersion relations in the forward limit $t =0$ does not work. As explained in \cite{Swamp}, this problem can be bypassed by working in the small-impact-parameter limit instead. We explain that this issue can be avoided by using the parametric dispersion relation, as the graviton pole contribution can be regulated by choosing a suitable non-zero value for the parameter, and thus, the partial wave expansions of all the Wilson coefficients converge. To demonstrate this, we derive bounds on the space of full crossing symmetric and meromorphic 2-2 scattering amplitudes with a graviton pole (see Fig. \ref{fig:g2g3plot}.)  — for example, the tree-level 2-2 dilaton amplitude in type II string theory belongs to this class.

Finally, we demonstrate that our formalism also allows us to derive series representations for multi-variable generalisations of the Veneziano and Virasoro–Shapiro amplitudes, which make the poles in all variables manifest. We expect that our multi-variable dispersive formula applies more generally. This represents a crucial first step toward a dispersive representation of $n$-particle amplitudes. Several technical developments remain before this goal can be fully realised, and we outline these in the text. Our preliminary results are intended to serve as a foundation for advancing this program.

The paper is organised as follows. In sections (\ref{sec:2ch}) and (\ref{sec:3ch}), we derive the two-channel and three-channel stringy dispersion relations starting with Cauchy's residue formula. In section (\ref{sec:examples}), we consider some examples of amplitudes that can be expressed using our dispersion relations. In section (\ref{sec:string}), we consider applications to string amplitudes and for bootstrapping weakly-coupled gravitational EFTs. In section (\ref{sec:npoint}), we outline the steps towards the $n$-particle dispersion relation. We conclude in section (\ref{sec:discussion}). An appendix at the end supplements the main text.

\section{The two-channel symmetric dispersion relation}
\label{sec:2ch}
Consider a 2-2 scattering amplitude $\mathcal{M}(s,t)$ that is symmetric under $s \leftrightarrow t$ exchange and satisfies $\displaystyle \lim_{|s| \to \infty}\left|\mathcal{M}\left(s,t\right)\right| = 0$ for some domain $\mathcal{D}$ in $t$. We assume that all the poles and branch cuts of the amplitude in the $s$-variable lie on a contour $\mathcal{C}_1$. Then the amplitude can be expressed via the following one-parameter, $s\leftrightarrow t$ symmetric dispersion relation, which converges for $-\lambda \in \mathcal{D}$. 
\begin{equation}
\label{2chParCSDR1}
    \mathcal{M}(s,t) = \frac{1}{\pi}\int_{\mathcal{C}_1} d\sigma~\mathcal{H}(s,t,\lambda) \mathcal{A}^{(s)}\left(\sigma,\frac{\left(s+\lambda\right)\left(t+\lambda\right)}{\sigma+\lambda}-\lambda\right)\,.
\end{equation}
where the kernel ${\mathcal H}(s,t,\lambda)$ is 
\begin{equation}
\label{KerCS}
{\mathcal H}^{(\l)}(\s,s,t)=\left(\frac{1}{\sigma-s}+\frac{1}{\sigma-t} -\frac{1}{\sigma+\lambda}\right)\,.
\end{equation}
and 
\begin{equation}
    \mathcal{A}^{(s)}\left(s,t\right) = \lim_{\epsilon \to 0^+} \frac{1}{2 i}\left(\mathcal{M}(s+i\epsilon,t) - \mathcal{M}(s-i\epsilon,t)\right)
\end{equation}
is the $s$-channel discontinuity of the amplitude across the contour $\mathcal{C}_1$. The dispersion relation is independent of the parameter $\lambda$. Further, when we plug in $\lambda = -t$, we recover the fixed-$t$ dispersion relation, while $\lambda = -s$ gives the fixed-$s$. Thus, we see that the parameter precisely plays the role of the ``elasticity" parameter of the string worldsheet mentioned before (see Fig. (\ref{fig:sdisp})), which enables us to deform between and equate the $s$-channel and $t$-channel representations. This exhibits the ``worldsheet duality" in string theory very naturally, prompting the name ``stringy" for such a dispersion relation. Let us now see how to derive it.

\subsection{Derivation} 
Assume that the amplitude satisfies $\displaystyle \lim_{|s| \to \infty}\left|\mathcal{M}\left(s,t\right)\right| = 0$ in some domain $\mathcal{D}$ of $t$. Then the following contour integral along $\Gamma$, a large circle at infinity, vanishes for $t \in \mathcal{D}$. 
\begin{equation}
\label{disp_fixed_t1}
	\frac{1}{2\pi i}\int_{\Gamma}\mathrm{d}\sigma \frac{1}{\sigma -s}\mathcal{M}\left(\sigma,t\right) = 0
\end{equation}
Shrinking the contour, we get a contribution from the pole at $\sigma = s$ which is $\mathcal{M}(s,t)$ and another contribution from the singularities of the amplitude, given in terms of the $s$-channel discontinuity $\mathcal{A}^{(s)}(s,t)$ along the contour $\mathcal{C}_1$. This leads to the fixed-$t$ dispersion relation. 
\begin{equation}
\label{disp_fixed_t2}
    \mathcal{M}(s,t) = \frac{1}{\pi}\int_{\mathcal{C}_1} d \sigma \frac{1}{\sigma -s} \mathcal{A}(\sigma,t)
\end{equation}
However, since the result in \eqref{disp_fixed_t1} only depends on the large $s$ behaviour of the amplitude, we can actually do a more general analysis. We can replace $\mathcal{M}\left(\sigma,t\right)$ by $\mathcal{M}\left(\hat{s}_1(\sigma,s,t), \hat{s}_2^{(\lambda)}(\sigma,s,t)\right)$ and as long as when $\sigma \to \infty$, $\hat{s}_1(\sigma,s,t) \to \infty$ and $ \hat{s}_2^{(\lambda)}(\sigma,s,t) \to -\lambda$ for some $-\lambda \in \mathcal{D}$, $\mathcal{M}\left(\hat{s}_1(\sigma,s,t), \hat{s}_2^{(\lambda)}(\sigma,s,t)\right)$ still vanishes at large $\sigma$. In general, choosing arbitrary $\hat{s}_1, \hat{s}_2$ functions can map the singularities along the contour $\mathcal{C}_1$ to very complicated contours. In order to have the simplest analytic structure in the $\sigma$ plane, we will restrict to functions that map straight line contours to straight line contours and satisfy the conditions we require. The only allowed class of mappings are affine linear transformations for $\hat{s}_1$ and fractional linear transformations for $\hat{s}_2$. Therefore, we consider
\begin{equation}
    \hat{s}_1(\sigma,s,t) = a \sigma + b, \quad \hat{s}_2^{\lambda}(\sigma,s,t) = \frac{-\lambda \sigma + b}{\sigma + d}
\end{equation}
Now our goal is to obtain an $s \leftrightarrow t$ symmetric dispersion relation. So we further demand 
\begin{equation}
    \hat{s}_1(\sigma,s,t)|_{\sigma = s,t}  = s ,t,  \quad \hat{s}_2^{(\lambda)}(\sigma,s,t)|_{\sigma = s,t} = t,s
\end{equation}
This uniquely fixes $\hat{s}_1(\sigma,s,t) = \sigma$ and $\hat{s}_2^{(\lambda)}(\sigma,s,t) = \frac{(s+ \lambda)(t+ \lambda)}{\sigma + \lambda} - \lambda$. We see that $\lambda$ is left over as a free parameter. Now let us explain the form of the kernel in \eqref{KerCS}. The first two terms are obvious from the requirement of crossing symmetry. The third term follows from the fact that $\hat{s}_2^{(\lambda)}(\sigma,s,t)|= t,s$ when $\lambda = -s, -t$. This tells us that the parametric dispersive representation must turn into the fixed-$t,s$ dispersion relations in the limits when $\lambda = -s, -t$, and therefore, we must necessarily have the third term in the kernel to cancel the poles in $s,t$-channels when $\lambda = -s,-t$.
\begin{equation}
\label{cont-disp}
	\frac{1}{2\pi i}\int_{\Gamma}\mathrm{d}\sigma{\mathcal H}^{(\lambda)}(\sigma,s,t)\mathcal{M}\left(\sigma,\hat{s}_2^{(\lambda)}(\sigma,s,t)\right) = 0
\end{equation}
where $\Gamma$ is a large circle at infinity. From here, the parametric dispersion relation in \eqref{2chParCSDR1} follows similarly to the fixed-$t$ case by shrinking the contour and picking up the singularities from the kernel and the amplitude. Note that while it may seem like there are new singularities being introduced at $\sigma + \lambda =0$ but this is avoided because $\mathcal{M}\left(\sigma, \frac{(s+ \lambda)(t+ \lambda)}{\sigma + \lambda} - \lambda\right)\Big|_{\sigma = -\lambda} =  \mathcal{M}\left(-\lambda, \infty\right) = \mathcal{M}\left(\infty, -\lambda\right) = 0$ due to our assumed large $s$ behavior. \\
While the above method is nice and intuitive, we will now present a different way to derive the parametric dispersion relation which rigorously shows that the form of $\hat{s}_2^{(\lambda)}(\sigma,s,t)$ has to be as we saw above once we fix the kernel to be of the form in \eqref{KerCS}. This method also generalises easily to the case of three-channel crossing symmetric amplitudes, which we discuss later. Let us treat $\hat{s}_2^{(\lambda)}(\sigma,s,t)$ as an arbitrary, as-of-now undetermined function We only assume for now that $\displaystyle \lim_{|\sigma| \to \infty}\hat{s}_2^{(\lambda)}(\sigma,s,t) \in \mathcal{D}$. Then
\begin{equation}
\begin{split}
 \lim_{|\sigma| \to \infty} \left|\mathcal{M}\left(\sigma,\hat{s}_2^{(\lambda)}(\sigma,s,t)\right)\right| =0 \,.
 \end{split}
\end{equation}
and we can write \eqref{cont-disp}.  If the amplitude grows faster at large $\sigma$ than we have assumed, we can introduce additional counter-terms/subtraction-terms in the integrand to ensure that the integral vanishes. We discuss such cases in the next section.
Now, let us contract the contour. We pick up singularities from both the kernel and the amplitude. The contributions from the amplitude lead to the following. 
\begin{eqnarray}
\label{AmpCont2ch}
 -\frac{1}{\pi}\int_{\mathcal{C}_{1}}\mathrm{d}\sigma\, \kernel\mathcal{A}^{(s)}\left(\sigma,\hat{s}_2^{(\lambda)}(\sigma,s,t)\right) 
  - \frac{1}{\pi}\int_{\mathcal{C}_{2}}\mathrm{d}\sigma\, \kernel\mathcal{A}^{(t)}\left(\sigma,\hat{s}_2^{(\lambda)}(\sigma,s,t)\right),  
\end{eqnarray}
where $\mathcal{A}^{(s)}(s,t)$ and $\mathcal{A}^{(t)}(s,t)$  are the discontinuities of the amplitude in the $s$-channel and $t$-channel respectively along the contour $\mathcal{C}_1$. Due to $s \leftrightarrow t$ symmetry of the amplitude, $\mathcal{A}^{(t)}(s,t) = \mathcal{A}^{(s)}(t,s)$. Crossing symmetry dictates that $\hat{s}_2^{(\lambda)}(\sigma,s,t)$ must map $\mathcal{C}_2$ to $\mathcal{C}_1$.  We now demand that the two terms in \eqref{AmpCont2ch} be equal. This can be achieved by doing a coordinate transformation $\sigma=\sigma(\tau)$ in the second term as follows
\begin{equation}
 \frac{1}{\pi}\int\mathrm{d}\tau\frac{d\sigma}{d\tau}\mathcal{H}^{(\lambda)}(\sigma(\tau),s,t)\mathcal{A}^{(t)}(\sigma(\tau), \hat{s}_2^{(\lambda)}(\sigma(\tau),s,t))=\frac{1}{\pi}\int\mathrm{d}\tau\frac{d\sigma}{d\tau}\mathcal{H}^{(\lambda)}(\sigma(\tau),s,t)\mathcal{A}^{(s)}(\hat{s}_2^{(\lambda)}(\sigma(\tau),s,t),\sigma(\tau))\,,
\end{equation}
and demanding that the following conditions be met
\begin{equation}
\label{conds}
\frac{d\sigma}{d\tau}\mathcal{H}^{(\lambda)}(\sigma(\tau),s,t)=\mathcal{H}^{(\lambda)}(\tau,s,t)\,,\qquad \hat{s}_2^{(\lambda)}(\sigma(\tau),s,t)=\tau\,,\quad \sigma(\tau)=\hat{s}_2^{(\lambda)}(\tau,s,t)\,.    
\end{equation}
The first condition leads to 
\begin{equation}
\frac{d\sigma}{d\tau}\frac{d}{d\sigma}\ln \frac{(\sigma-s)(\sigma-t)}{(\lambda+\sigma)}=\frac{d}{d\tau}\ln \frac{(\tau-s)(\tau-t)}{(\lambda+\tau)}\,,    
\end{equation}
giving 
\begin{equation}
 \frac{(\sigma-s)(\sigma-t)}{(\lambda+\sigma)}=c \frac{(\tau-s)(\tau-t)}{(\lambda+\tau)}\,,   
\end{equation}
where $c$ is an integration constant. The second condition fixes $c=1$ and also maps $\mathcal{C}_2$ to $\mathcal{C}_1$. We thus get a quadratic equation for $\tau$ with two solutions:
\begin{equation}
 \tau =\sigma\,,\qquad \tau=\frac{\left(s+\lambda\right)\left(t+\lambda\right)}{\sigma+\lambda}-\lambda\implies \sigma=\frac{\left(s+\lambda\right)\left(t+\lambda\right)}{\tau+\lambda}-\lambda\,.  
 \end{equation} 
The first solution has no $s,t$ dependence and therefore we discard it. The second solution leads to precisely the form of $\hat{s}_2^{(\lambda)}(\sigma,s,t)$ that we found before. Using the second solution, the contribution from the singularities of the amplitude is 
\begin{equation}
-\frac{2}{\pi}\int_{\mathcal{C}_{1}}\mathrm{d}\sigma\, \kernel\mathcal{A}^{(s)}\left(\sigma,\frac{(s+\lambda)(t+\lambda)}{\sigma+\lambda}-\lambda\right) \,.
\end{equation}
Now, we come to the singularities that arise from the kernel $\mathcal{H}(s,t,\lambda)$. As explained before, the apparent singular term at $\sigma = -\lambda$ is actually not singular due to our assumed large $\sigma$ behaviour. Therefore, 
\begin{equation}
    \underset{\sigma = -\lambda}{\text{Res}} ~\frac{1}{\sigma+\lambda}\mathcal{M}\left(\sigma,\frac{(s+\lambda)(t+\lambda)}{\sigma+\lambda}-\lambda \right) = 0
\end{equation}
Therefore, we simply pick up the residues from the poles at $\sigma = s, t$, giving $\mathcal{M}(s,t) + \mathcal{M}(t,s) = 2\mathcal{M}(s,t)$. Collecting the contributions from all the singularities, we arrive at
\begin{equation}
\label{2chParCSDR2}
    \mathcal{M}(s,t) = \frac{1}{\pi}\int_{\mathcal{C}_{1}}\mathrm{d}\sigma\, \kernel\mathcal{A}^{(s)}\left(\sigma,\frac{\left(s+\lambda\right)\left(t+\lambda\right)}{\sigma+\lambda}-\lambda\right) \,.
\end{equation}
This is the two-channel symmetric dispersion relation with a one-parameter ambiguity given in \eqref{2chParCSDR1}, which converges for all $s,t$ when $-\lambda \in \mathcal{D}$.
 \begin{figure}[H]
    \centering
    \includegraphics[width= 0.6\linewidth]{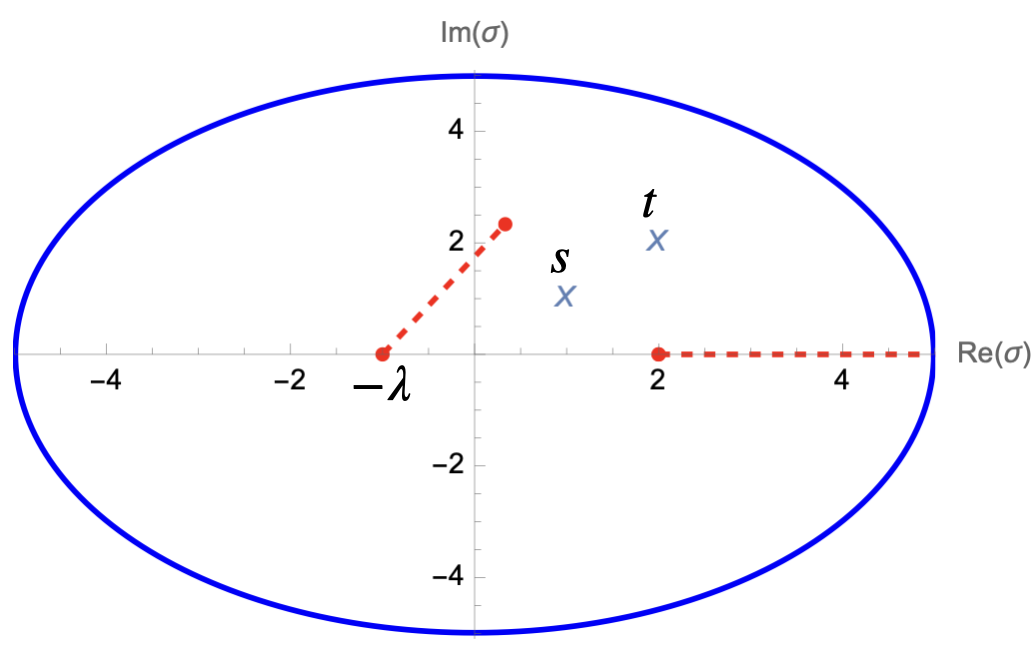}
    \caption{A typical depiction of the singularities in the integrand in the complex $\sigma$ plane. We consider the amplitude $\frac{1}{\sqrt{2-s}} + \frac{1}{\sqrt{2-t}}$. Then $\mathcal{C}_1$ is the contour $\sigma \ge 2$ and $\mathcal{C}_2$ is the contour $\hat{s}_2^{(\lambda)} (\sigma, s,t) \ge 2$. In the plot, we choose $s=1+i, \: t=2+2i, \: \lambda=1$ to determine $\mathcal{C}_2$. It is the line joining $(1/3,7/3)$ and $(-1,0)$.} 
    \label{fig:mdisp}
\end{figure}

\subsection{Higher subtractions}
If an amplitude grows faster than $o\left(s^{0}\right)$ in the large $s$ limit, then we need subtractions/counter terms in the dispersion relation. There is no unique choice of adding such terms, and here we present two forms of subtracted dispersion relations. 

\subsubsection{First method}
    Let us assume the amplitude satisfies $\displaystyle \lim_{|s| \to \infty} |\mathcal{M}(s,t)/s^N|  = 0$. Then one can consider the following integrand with additional subtraction terms to ensure the integral over a large circle at infinity vanishes.
    \begin{eqnarray}  \label{sub-int}  &&\mathcal{H}^{(\lambda)}\left(\sigma,s,t\right)\mathcal{M}\left(\sigma,\frac{\left(s+\lambda\right)\left(t+\lambda\right)}{\lambda+\sigma}-\lambda\right)\nonumber\\
    &&- \sum_{p =0}^{N-1}\left(s+\lambda\right)^{p}\left(t+\lambda\right)^{p}\left[\frac{\partial^{p}}{\partial s^{p}}\frac{\partial^{p}}{\partial t^{p}}\mathcal{H}^{(\lambda)}\left(\sigma,s,t\right)\mathcal{M}\left(\sigma,\frac{\left(s+\lambda\right)\left(t+\lambda\right)}{\lambda+\sigma}-\lambda\right)\right]_{s=-\lambda, t=-\lambda}
    \end{eqnarray}
    Then we can write
    \begin{equation}
        \int_{\Gamma}\mathrm{d}\sigma \left[\eqref{sub-int}\right] = 0.
    \end{equation}
     More generally, we can consider a function $f\left(x,y;\sigma\right)$ that satisfies
    \begin{equation}
        \lim_{|\sigma|\rightarrow\infty}\biggl\{f\left(x,y;\sigma\right)-f\left(x_{0},y_{0};\sigma\right)\biggr\} =0.
    \end{equation}
    Then Taylor expanding $f$ around $x_{0}$ and $y_{0}$, we can also demand that 
    \begin{equation}
        \lim_{|\sigma|\rightarrow\infty}\biggl\{f\left(x,y;\sigma\right)-\sum_{p,q}\left(x-x_{0}\right)^{p}\left(y-y_{0}\right)^{q}\frac{\partial^{p}}{\partial x^{p}}\frac{\partial^{q}}{\partial y^{q}}f\left(x_{0},y_{0};\sigma\right)\biggr\} =0.
    \end{equation}
    The advantage of subtracting higher-order terms, coming from the Taylor series expansion, is that terms that are more divergent in $\sigma$ get removed. 
    Let us consider the following function of $\sigma$,
    \begin{equation}\label{func}
       f\left(\sigma;s,t\right) =\left(\frac{1}{s-\sigma}+\frac{1}{t-\sigma}+\frac{1}{\lambda+\sigma}\right)\widetilde{\mathcal{M}}\left(\frac{\left(s+\lambda\right)\left(t+\lambda\right)}{\lambda+\sigma}-\lambda\right),
    \end{equation}
    where we denote $\mathcal{M}\left(\sigma,\frac{\left(s+\lambda\right)\left(t+\lambda\right)}{\sigma+\lambda}-\lambda\right)=\widetilde{\mathcal{M}}\left(\frac{\left(s+\lambda\right)\left(t+\lambda\right)}{\lambda+\sigma}-\lambda\right)$. Expanding $f\left(\sigma; s, t\right)$ around $s=-\lambda$ and $t=-\lambda$, we obtain
    \begin{equation}\label{exp1}
      f\approx  -\frac{1}{\lambda+\sigma}\widetilde{\mathcal{M}}\left(-\lambda\right) - \frac{1}{\left(\lambda+\sigma\right)^{2}}\biggl\{\left(s+t+2\lambda\right)\widetilde{\mathcal{M}}\left(-\lambda\right)+\left(s+\lambda\right)\left(t+\lambda\right)\widetilde{\mathcal{M}}'\left(-\lambda\right)\biggr\} + \mathcal{O}\left(s^{2}, t^{2}\right).
    \end{equation}
    Here, the derivative is taken with respect to $\sigma$. Again, if we expand $f\left(\sigma;s,t\right)$ around $\sigma=\infty$, we get
    \begin{equation}\label{exp2}
        f\approx -\frac{1}{\sigma}\widetilde{\mathcal{M}}\left(-\lambda\right) - \frac{1}{\sigma^{2}}\biggl\{\left(s+t+2\lambda\right)\widetilde{\mathcal{M}}\left(-\lambda\right)+\left(s+\lambda\right)\left(t+\lambda\right)\widetilde{\mathcal{M}}'\left(-\lambda\right)\biggr\} + \mathcal{O}\left(\frac{1}{\sigma^{3}}\right).
    \end{equation}
    From the above expressions, we notice that \eqref{exp1} in the limit $\sigma\rightarrow\infty$ produces \eqref{exp2}. This feature holds for the higher-order terms in the expansions.  Therefore, subtracting the series expansion around $s=-\lambda$ and $t=-\lambda$ given in \eqref{exp1} from \eqref{func} is equivalent to adding counterterms such that divergences are removed at $\sigma=\infty$.
    
    For the purpose of illustration, we can consider the following example,
    \begin{equation}
        \mathcal{M}\left(s,t\right) = st\left(\frac{1}{s-1}+\frac{1}{t-1}\right).
    \end{equation}
    In this case, the first term in eq.(\ref{sub-int}) 
    behaves as $-\frac{\lambda}{\lambda+1}+\frac{st+\lambda-\left(s+t-2\lambda\right)\lambda^{2}}{\left(1+\lambda\right)^{2}\sigma}$ in the large $\sigma$ limit. Let us expand the function about 
    $s=-\lambda$, $t=-\lambda$. The leading term of the series is $\frac{\lambda\left(2+\lambda-\sigma\right)\sigma}{\left(1+\lambda\right)\left(\lambda+\sigma\right)\left(\sigma-1\right)}$ which in the large $\sigma$ limit behaves as $-\frac{\lambda}{1+\lambda}+\frac{\lambda\left(1+2\lambda\right)}{\left(1+\lambda\right)\sigma}$. By adding this as counter term, although the $\sigma^{0}$ term is cancelled, we are still left with $\sigma^{-1}$ term which contributes to a pole at infinity. This problem is cured by keeping terms to $\mathcal{O}\left(\left(s+\lambda\right)\left(t+\lambda\right)\right)$ in the series expansion. In general, the strategy is to include counter terms such that the large $\sigma$ fall-off is faster than $\sigma^{-1}$.

The $N$-subtracted dispersion relation takes the form
\begin{eqnarray}\label{HS-method1}
    \mathcal{M}\left(s,t\right) &=&     \sum_{p=0}^{N-1}\left(s+\lambda\right)^{p}\left(t+\lambda\right)^{p}\left[\frac{\partial^{p}}{\partial s^{p}}\frac{\partial^{p}}{\partial t^{p}}\mathcal{M}\left(s,t\right)\right]_{s=-\lambda,t=-\lambda} \nonumber\\
    &&-\frac{1}{\pi}\int_{\mathcal{C}_{1}}\mathrm{d}\sigma\Biggl\{\mathcal{H}^{(\lambda)}\left(\sigma,s,t\right)\mathcal{A}^{(s)}\left(\sigma,\frac{\left(s+\lambda\right)\left(t+\lambda\right)}{\sigma+\lambda}-\lambda\right)\nonumber\\
        && -\sum_{p=0}^{N-1}\left(s+\lambda\right)^{p}\left(t+\lambda\right)^{p}\left[\frac{\partial^{p}}{\partial s^{p}}\frac{\partial^{p}}{\partial t^{p}}\mathcal{H}^{(\lambda)}\left(\sigma,s,t\right)\mathcal{A}^{(s)}\left(\sigma,\frac{\left(s+\lambda\right)\left(t+\lambda\right)}{\lambda+\sigma}-\lambda\right)\right]_{s=-\lambda,t=-\lambda} \Biggr\}\nonumber\\
\end{eqnarray}
The steps to obtain the above expression have been delineated in the Appendix \ref{app:HS}. Putting $N = 1$, we get the once-subtracted, parametric CSDR  which was obtained in \cite{appl4}. Explicitly, it is given as
\begin{equation}
\label{Disp2chSub}
\begin{split}
 \mathcal{M}\left(s,t\right) = \mathcal{M}\left(-\lambda,-\lambda\right) - \frac{1}{\pi}\int_{\mathcal{C}_{1}}\mathrm{d}\sigma\Biggl\{\left(\frac{1}{s-\sigma}+\frac{1}{t-\sigma}+\frac{1}{\lambda+\sigma}\right)&\mathcal{A}^{(s)}\left(\sigma,\frac{\left(s+\lambda\right)\left(t+\lambda\right)}{\lambda+\sigma}-\lambda\right) \\
 &~~~~~~~~~+ \frac{1}{\lambda+\sigma}\mathcal{A}^{(s)}\left(\sigma,-\lambda\right)\Biggr\} \,.     
\end{split}
\end{equation} 

\subsubsection{Second method}
Here, we first construct a two-channel symmetric function from the amplitude,
\begin{equation}
    \widehat{\mathcal{M}}\left(s,t\right)=\frac{\mathcal{M}\left(s,t\right)}{\left(s-\alpha\right)^{N}\left(t-\alpha\right)^{N}},
\end{equation}
where $N$ is an integer and $\alpha$ is arbitrary. $N$ is chosen such that 
\begin{equation}
\mathcal{H}^{(\lambda)}\left(\sigma,s,t\right)\widehat{\mathcal{M}}\left(\sigma,\frac{\left(s+\lambda\right)\left(t+\lambda\right)}{\lambda+\sigma}-\lambda\right)
\end{equation}
falls off faster than $1/\sigma$ as $\sigma\rightarrow\infty$. Then we can apply the unsubtracted dispersion relation to $\widehat{\mathcal{M}}\left(s,t\right)$.  

Consider $\mathcal{M}\left(s,t\right) = st\left(\frac{1}{s-1}+\frac{1}{t-1}\right)$. If we divide this amplitude by $\left(s-\alpha\right)^{2}\left(t-\alpha\right)^{2}$ we get the correct behaviour at infinity.  The $s$-channel discontinuities of $\widehat{\mathcal{M}}\left(s,t\right)$ have a simple pole at $s=1$ and a double pole at $s=\alpha$. It can be checked that 
\begin{equation}
    \mathcal{M}\left(s,t\right) = \left(s-\alpha\right)^{2}\left(t-\alpha\right)^{2} \sum_{\sigma=1, \alpha}\mathrm{Res}\left[\mathcal{H}^{(\lambda)}\left(\sigma,s,t\right)\widehat{\mathcal{M}}\left(\sigma,\frac{\left(s+\lambda\right)\left(t+\lambda\right)}{\lambda+\sigma}-\lambda\right)\right].
\end{equation}
Note that although individual residues depend on $\lambda$, the final answer obtained after adding them is independent of $\lambda$.

In general, $\mathcal{M}\left(s,t\right)$ can be multi-valued functions of $s$ and $t$, and have branch-cut singularities. We can choose $\alpha$ so that it does not lie on the cut along the discontinuities of $\mathcal{M}\left(s,t\right)$. In that case, the amplitude can be given by 
\begin{eqnarray}
    \mathcal{M}\left(s,t\right) &=&  \left(s-\alpha\right)^{N}\left(t-\alpha\right)^{N}\Biggl\{-\frac{1}{\pi}\int_{\mathcal{C}}\mathrm{d}\sigma\; \mathcal{H}^{(\lambda)}\left(\sigma, s, t\right)\frac{\mathcal{A}^{(s)}\left(\sigma,\frac{\left(s+\lambda\right)\left(t+\lambda\right)}{\lambda+\sigma}-\lambda\right)}{\left(\sigma-\alpha\right)^{N}\left(\frac{\left(s+\lambda\right)\left(t+\lambda\right)}{\lambda+\sigma}-\lambda-\alpha\right)^{N}}\nonumber\\
    && \hspace{1cm} + \frac{1}{2\pi i}\oint_{\{\sigma\rightarrow\alpha\}}\mathrm{d}\sigma\; \mathcal{H}^{(\lambda)}\left(\sigma,s,t\right)\widehat{\mathcal{M}}\left(\sigma,\frac{\left(s+\lambda\right)\left(t+\lambda\right)}{\lambda+\sigma}-\lambda\right)\Biggr\},
\end{eqnarray}
where $\mathcal{C}$ represents the contour along the cut. 

We take another example: $\mathcal{M}\left(s,t\right) = st\left(\frac{1}{\left(1-s\right)^{\frac{1}{3}}} + \frac{1}{\left(1-t\right)^{\frac{1}{3}}}\right)$. The $s$-channel discontinuity of this amplitude is $\mathcal{A}^{(s)}\left(s,t\right) = \frac{st}{\left(s-1\right)^{\frac{1}{3}}}\sin\left(\frac{\pi}{3}\right)$, with $s>1$. The following table illustrates how $\lambda$ moves the contributions around while making the final answer invariant.
\begin{table}[H]
\begin{center}
\begin{tabular}{c|c|c|c}
$\lambda$ & $b_{1}$ & $b_{2}$ & $b_{2}-b_{1} = \widehat{\mathcal{M}}$\\
\hline
 $-2.52381+i 1.85294$  & $-0.00665292+i 3.16006 $ & $-2.51847+i 1.58393$  & $-2.51181-i 1.57613$  \\
 $3.20588 -i 1.04167$  & $-0.0249031-i 0.0463166$  & $-2.53672-i 1.62245$  & $-2.51181-i 1.57613$  \\
 $3.22222 -i 1.91489$  & $-0.0174898-i 0.0387658$  & $-2.5293- i 1.6149$  & $-2.51181-i 1.57613$  \\
 $-0.0810811+i 0.0227273$  & $0.0209022 +i 3.11311$  & $-2.49091+i 1.53698$  & $-2.51181-i 1.57613$  \\
 $1.55 +i 2.49315$  & $-0.00338277+i 3.18388$  & $-2.5152+i 1.60775$  & $-2.51181-i 1.57613$  \\
\end{tabular}
\caption{We evaluate $\frac{\mathcal{M}\left(s,t\right)}{\left(s-\alpha\right)\left(t-\alpha\right)}$ with $s=2.5+ i 10^{-10}, t=-1.2+i 10^{-10}, \alpha=-1.5+i 10^{-10}$ for different values of $\lambda$. $b_{1}$ is the contribution from the branch cut and $b_{2}$ is the contribution from the pole at $\sigma=\alpha$. }
\end{center}
\end{table}

\section{The three-channel symmetric dispersion relation}
\label{sec:3ch}
Let us now discuss amplitudes which have three-channel symmetry, i.e., invariance under permutations of $s,t,u$. If the external particles have a mass $m$, then $s + t  + u = 4m^2$. For convenience, we will work with shifted Mandelstam variables, defined as
\begin{equation}
    s_1 = s -\frac{4m^2}{3}, \quad s_1 = t -\frac{4m^2}{3}, \quad s_3 = u -\frac{4m^2}{3}, \qquad s_1 + s_2 + s_3 = 0\,.
\end{equation}
We assume that the amplitude $\mathcal{M}(s_1,s_2,s_3)$ satisfies $\displaystyle
  \lim_{|s_1|\to \infty} \left|\mathcal{M}\left(s_1,s_2,s_3\right)\right| = 0$ for $s_2 \in \mathcal{D}$ and that all the poles and branch cuts in the $s$-variable lie on a contour $\mathcal{C}_1$. Then it can be expressed via the following one-parameter, crossing symmetric dispersion relation 
\begin{equation}
\label{3ch_CSDisp_final1}
\begin{split}
    &\mathcal{M}(s_1,s_2,s_3) = 
   \frac{1}{\pi}\int_{\mathcal{C}_1}d \sigma \mathcal{H}^{(\lambda)}(\sigma, s_1,s_2,s_3)  \mathcal{A}^{(s_1)}\left(\sigma,\hat{s}_2^{(\lambda)}(\sigma,s_1,s_2,s_3), -\sigma -\hat{s}_2^{(\lambda)}(\sigma,s_1,s_2,s_3)\right)\,,
\end{split}
\end{equation}
for all $s_1,s_2$ and for $-\lambda \in \mathcal{D}$. Here, $\hat{s}_3^{(\lambda)}(\sigma,s_1,s_2,s_3) = - \sigma- \hat{s}_2^{(\lambda)}(\sigma,s_1,s_2,s_3)$, and
\begin{equation}
\begin{split}
     \hat{s}_2^{(\lambda)}(\sigma,s_1,s_2,s_3) &=  -\frac{\sigma}{2} \pm \sqrt{\frac{y+\lambda x}{\sigma+\lambda}+\frac{ \sigma^3- 3\lambda \sigma^2}{4(\sigma+\lambda)}}\,, \\
     y = -s_1 s_2 s_3, &\quad x = -(s_1 s_2 + s_2s_3 + s_3s_1)\,.
     \end{split}
\end{equation}
The kernel is given as
\begin{equation}
 \mathcal{H}^{(\lambda)}(\sigma, s_1,s_2,s_3) = \frac{1}{\sigma -s_1}+\frac{1}{\sigma-s_2}+\frac{1}{\sigma-s_3}-\frac{1}{\sigma+\lambda}\,,   
\end{equation}
and $\mathcal{A}^{(s_1)}\left(s_1,s_2, s_3\right)$ is the discontinuity of the amplitude across $\mathcal{C}_1$. The dispersion relation is independent of $\lambda$, and substituting $\lambda = -s_1,-s_2,-s_3$ leads to the fixed-$s_1,s_2,s_3$ dispersion relations, revealing the "stringy" nature of this dispersion relation, similar to the two-channel case.

\subsection{Derivation}
The derivation is similar to the two-channel case. We assume $\displaystyle
  \lim_{|s_1|\to \infty} \left|\mathcal{M}\left(s_1,s_2,s_3\right)\right| = 0$ for $s_2 \in \mathcal{D}$. This implies that
  \begin{equation}
  \lim_{|\sigma| \to \infty} \left|\mathcal{M}\left(\sigma,\hat{s}_2^{(\lambda)}(\sigma,s_1,s_2,s_3),\hat{s}_3^{(\lambda)}(\sigma,s_1,s_2,s_3)\right)\right| = 0\,.
\end{equation}
 when $\displaystyle \lim_{|\sigma| \to \infty} \hat{s}_2^{(\lambda)}(\sigma,s_1,s_2,s_3)\in \mathcal{D}$. $\hat{s}_2^{(\lambda)}(\sigma,s_1,s_2,s_3)$ is an as-of-yet undetermined function. Now, the following contour integral along a large circle at infinity vanishes.
\begin{equation}
\label{3chDispStart}
\begin{split}
    \frac{1}{2\pi i} \int_{\Gamma} d\sigma \mathcal{H}^{(\lambda)}(\sigma, s_1,s_2,s_3) \mathcal{M}(\sigma, \hat{s}_2^{(\lambda)}(\sigma,s_1,s_2,s_3), \hat{s}_3^{(\lambda)}(\sigma,s_1,s_2,s_3)) =0
\end{split}
\end{equation}
Again, the first three terms in the kernel $\mathcal{H}^{(\lambda)}(\sigma, s_1,s_2,s_3)$ are naturally motivated if we want to obtain a three-channel symmetric dispersion relation, while the motivation for the last $\lambda$-dependent term is that it allows us to get to the fixed-$s,t,u$ dispersion relations by choosing $\lambda = -s_1,-s_2,-s_3$ respectively.  From now on, for convenience, we will use the shorthand $\hat{s}_2^{(\lambda)}(\sigma,s_i) \equiv \hat{s}_2^{(\lambda)}(\sigma,s_1,s_2,s_3)$ and similarly for $\hat{s}_3$. Contracting the contour, we pick up singularities from the kernel as well as the amplitude. The latter gives
 \begin{equation}
 \label{3ch_first}
 \begin{split}
     & - \frac{1}{\pi}\int_{\mathcal{C}_1}d\sigma \mathcal{H}^{(\lambda)}(\sigma, s_1,s_2,s_3) \mathcal{A}^{(s_1)}\left(\sigma, \hat{s}_2^{(\lambda)}(\sigma,s_i),\hat{s}_3^{(\lambda)}(\sigma,s_i)\right)\\
     &- \frac{1}{\pi}\int_{\mathcal{C}_2}d\sigma \mathcal{H}^{(\lambda)}(\sigma, s_1,s_2,s_3) \mathcal{A}^{(s_2)}\left(\sigma, \hat{s}_2^{(\lambda)}(\sigma,s_i),\hat{s}_3^{(\lambda)}(\sigma,s_i)\right)\\
      &- \frac{1}{\pi}\int_{\mathcal{C}_3}d\sigma \mathcal{H}^{(\lambda)}(\sigma, s_1,s_2,s_3) \mathcal{A}^{(s_3)}\left(\sigma, \hat{s}_2^{(\lambda)}(\sigma,s_i),\hat{s}_3^{(\lambda)}(\sigma,s_i)\right)
\end{split}
\end{equation}
We want all three contributions above to be equal. We rewrite the second term as
\begin{equation}
\begin{split}
-\frac{1}{\pi} \int_{\mathcal{C}_2}d \hat{s}_2 \frac{d \sigma(\hat{s}_2)}{d \hat{s}_2} \mathcal{H}^{(\lambda)}(\sigma(\hat{s}_2)), s_1,s_2,s_3) \mathcal{A}^{(s_2)}(\sigma(\hat{s}_2), \hat{s}_2, \hat{s}_3(\hat{s}_2)) &\\ 
=  -\frac{1}{\pi} \int_{\mathcal{C}_1}d \hat{s}_2 \frac{d \sigma(\hat{s}_2)}{d \hat{s}_2} \mathcal{H}^{(\lambda)}(\sigma(\hat{s}_2)), s_1,s_2,s_3) \mathcal{A}^{(s_1)}(\hat{s}_2, \sigma(\hat{s}_2), \hat{s}_3(\hat{s}_2)) )
\end{split}
\end{equation}
where in the last line we used $\mathcal{A}^{(s_2)}(s_1, s_2,s_3) = \mathcal{A}^{(s_1)}(s_2, s_1,s_3)$ which follows from the crossing symmetry of the amplitude. Equality of the first two terms in \eqref{3ch_first} requires
    \begin{equation}
    \label{3ch_condition}
    \begin{split}
     \frac{d \sigma(\hat{s}_2)}{d \hat{s}_2} \mathcal{H}^{(\lambda)}(\sigma(\hat{s}_2)), s_1,s_2,s_3) &= \mathcal{H}^{(\lambda)}(\hat{s}_2, s_1,s_2,s_3) \quad \text{and} \quad \hat{s}_2(\sigma)=\sigma(\hat{s}_2)
    \end{split}
\end{equation}
This first condition can be rewritten as
\begin{equation}
\frac{\partial}{\partial\hat{s}_2} \log\left[ \frac{(s_1-\sigma(\hat{s}_2))(s_2-\sigma(\hat{s}_2))(s_3-\sigma(\hat{s}_2))}{(\sigma(\hat{s}_2)+\lambda)}\right] = \frac{\partial}{\partial\hat{s}_2} \log\left[ \frac{(s_1-\hat{s}_2)(s_2-\hat{s}_2)(s_3-\hat{s}_2)}{(\hat{s}_2+\lambda)}\right]
\end{equation}
which implies
\begin{equation}
   \frac{(s_1-\sigma(\hat{s}_2))(s_2-\sigma(\hat{s}_2))(s_3-\sigma(\hat{s}_2))}{(\sigma(\hat{s}_2)+\lambda)} = c \frac{(s_1-\hat{s}_2)(s_2-\hat{s}_2)(s_3-\hat{s}_2)}{(\hat{s}_2+\lambda)}
\end{equation}
The second condition fixes $c = 1$. Solving the resulting equation gives
\begin{equation}
\label{3ch_allsols}
    \sigma(\hat{s}_2) =  \hat{s}_2 , \quad \sigma^{\pm}(\hat{s}_2) = -\frac{\hat{s}_2}{2} \pm \sqrt{\frac{y+\lambda x}{\hat{s}_2+\lambda}+\frac{ \hat{s}_2^3- 3\lambda \hat{s}_2^{2}}{4(\hat{s}_2+\lambda)}}\,.
\end{equation}
We will consider the $\sigma^{+}$ solution. This leads to 
\begin{equation}
    \hat{s}_2^{(\lambda)}(\sigma,s_i) =  -\frac{\sigma}{2} + \sqrt{\frac{y+\lambda x}{\sigma+\lambda}+\frac{ \sigma^3- 3\lambda \sigma^2}{4(\sigma+\lambda)}}\,,
\end{equation}
We still have to deal with the third term in \eqref{3ch_first}. It needs to satisfy conditions analogous to \eqref{3ch_condition} and, therefore, we get solutions analogous to \eqref{3ch_allsols}. It is easy to see that in this case, we pick the solution $\sigma^-(s_3)$ as that leads to 
\begin{equation}
    \hat{s}_3^{(\lambda)}(\sigma,s_i) =  -\frac{\sigma}{2} - \sqrt{\frac{y+\lambda x}{\sigma+\lambda}+\frac{ \sigma^3- 3\lambda \sigma}{4(\sigma+\lambda)}}\,,
\end{equation}
and naturally gives $\hat{s}_3^{(\lambda)}(\sigma,s_i)= -\sigma -  \hat{s}_2^{(\lambda)}(\sigma,s_i)$ as required. Therefore, the total contribution from the singularities of the amplitude is 
\begin{equation}
    - \frac{3}{\pi}\int_{\mathcal{C}_1}d\sigma \mathcal{H}^{(\lambda)}(\sigma, s_1,s_2,s_3) \mathcal{A}^{(s_1)}(\sigma,  \hat{s}_2^{(\lambda)}(\sigma,s_i), \hat{s}_3^{(\lambda)}(\sigma,s_i))\,.
\end{equation}
Now, coming to the contribution from singularities of the kernel. As in the two-channel case, our assumed large $\sigma$ behaviour means that the $1/(\sigma+\lambda)$ term gives no contribution. The residues from the poles at $\sigma = s_1,s_2,s_3$ give in total $3 \mathcal{M}(s_1,s_2,s_3)$.  Thus, combining the contributions of all the singularities, we arrive at
 \begin{equation}
 \label{3ch_CSDisp_final2}
     \mathcal{M}(s_1,s_2,s_3) =
      \frac{1}{\pi}\int_{\mathcal{C}_1}d\sigma \mathcal{H}^{(\lambda)}(\sigma, s_1,s_2,s_3) \mathcal{A}^{(s_1)}(\sigma, \hat{s}_2^{(\lambda)}(\sigma,s_i),\hat{s}_3^{\lambda}(\sigma,s_i))\,.
 \end{equation}
This is the parametric, three-channel, crossing symmetric dispersion relation given in \eqref{3ch_CSDisp_final1} which converges for all $s_1,s_2$ and for $-\lambda \in \mathcal{D}$.

\subsection{Higher subtractions}
Like the two-channel symmetric case, we require generalisations of \eqref{3ch_CSDisp_final1} for amplitudes that grow faster at large values of $s$. As an example, if $\displaystyle \lim_{|s_1|\rightarrow \infty} \mathcal{M}\left(s_{1}, s_{2}, s_{3}\right) =o\left(s_{1}^{2}\right)$, then we need the twice-subtracted version of \eqref{3ch_CSDisp_final1} given as follows,
\begin{eqnarray}
\label{TwiceSubCSDR}
    \mathcal{M}\left(s_{1}, s_{2}, s_{3}\right) = \mathcal{M}\left(-\lambda,-\lambda,2\lambda\right)- \frac{1}{\pi}\int_{\mathcal{C}_{1}}\mathrm{d}\sigma \Biggl\{&\mathcal{H}^{(\lambda)}(\sigma, s_1,s_2,s_3) \mathcal{A}^{(s_1)}(\sigma, \hat{s}_2^{(\lambda)}\left(\sigma,s_i),\hat{s}_3^{\lambda}(\sigma,s_i)\right)\nonumber\\
    & +\left(\frac{1}{\lambda+\sigma}-\frac{1}{2\lambda-\sigma}\right)\mathcal{A}^{(s_{1})}\left(\sigma,-\lambda,-\sigma+\lambda\right)\Biggr\}.
\end{eqnarray}
The case with $\lambda=0$ leads to the dispersion relation presented in \cite{song}. For higher-subtracted, fully crossing-symmetric dispersion relations, we follow the same procedure as outlined in \eqref{HS-method1}.

\section{Some examples}
\label{sec:examples}
\subsection{$\frac{1}{s t}$}
We begin with a simple but illustrative example--${\mathcal M}(s,t)=\frac{1}{st}$. Since the amplitude vanishes at large $s$, we can use the unsubtracted two-channel dispersion relation in \eqref{2chParCSDR1}. The absorptive part in the dispersion relation is proportional to $ \delta(\sigma)/t$. The kernel contributes 
\begin{equation}
    \frac{1}{s}+\frac{1}{t}+\frac{1}{\lambda}\,,
\end{equation}
while the absorptive part, after replacing $t\rightarrow (s+\lambda)(t+\lambda)/(\sigma+\lambda)-\lambda$ contributes
\begin{equation}
    \frac{1}{s+t+\frac{s t}{\lambda}}\,.
\end{equation}
This means that each piece in the kernel (e.g. the $1/s$ term) comes with a spurious pole at $s+t+ st/\lambda=0$ which cancels when all three terms in the kernel are added. Furthermore, $\lambda$ here is vital. In the previous incarnation of the dispersion relation \cite{SZ, song}, this $\lambda$ was absent, which required subtracting the massless poles by hand. This is no longer needed, as $\lambda$ serves as an IR regulator. 

\subsection{$\frac{1}{s}+\frac{1}{t}$}
The amplitude goes to a constant at large $s$ in this case, and so we use \eqref{Disp2chSub}. The absorptive part of the amplitude is trivial and is given by $\mathcal{A}^{(s)}\left(s,t\right)= -\pi\delta\left(s\right) \equiv\mathcal{A}^{(s)}\left(s\right)$.  Since this is independent of $t$, we can write
\begin{eqnarray}
    \mathcal{M}\left(s,t\right) & = & \mathcal{M}\left(-\lambda,-\lambda\right) - \frac{1}{\pi}\int\mathrm{d}\sigma \left(\frac{1}{s-\sigma}+\frac{1}{t-\sigma}+\frac{2}{\lambda+\sigma}\right)\mathcal{A}^{(s)}\left(\sigma\right).
\end{eqnarray}
$\mathcal{M}\left(-\lambda,-\lambda\right)$ cancels with the last term in the kernel, and we recover $\frac{1}{s}+\frac{1}{t}$.

\subsection{One-loop box diagram}
 \begin{figure}[H]
	\centering
	\begin{tikzpicture}
            \draw[thin] (-2,1)--(-1,0);
            \draw[thin] (2,1)--(1,0);
            \draw[thin] (-2,-3)--(-1,-2);
            \draw[thin] (1,-2)--(2,-3);
		\draw [line width = 0.1mm] (-1,0) -- (1,0);
		\draw[line width = 0.8mm] (-1,0) -- (-1,-2);
		\draw[line width=0.8mm] (1,0) -- (1,-2);
		\draw [line width=0.1mm] (-1,-2) --(1,-2);
	\end{tikzpicture}
    \caption{Thick lines denote massive states in the loop and massless legs are marked with thin lines}
\end{figure}
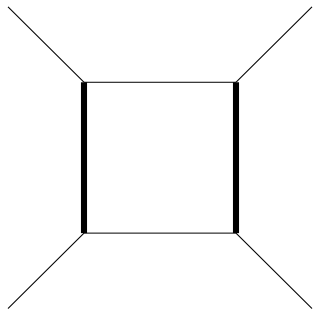
Dispersion relations can be used in the evaluation of Feynman diagrams \cite{Britto:2024mna, Bargiela:2024hgt}. The one-loop box diagram in four dimensions with two massive and two massless propagators takes the following form \cite{Smirnov:2004ym}

	\begin{equation}
		\mathcal{I}\left(s,t\right) = \frac{1}{t\sqrt{s\left(s-4\right)}}\left(\frac{1}{\epsilon}-\log\left(-t\right)\right)\log\left(1-\frac{s-\sqrt{s\left(s-4\right)}}{2}\right) + \left(s\leftrightarrow t\right) + \mathcal{O}\left(\epsilon\right)
	\end{equation}
    $\epsilon$ is the dimensional regulation parameter. This expression diverges at $s=0$ and $t=0$. In the full amplitude, there are vertex factors which multiply the above expression. Let us consider the them of the form $\left(st\right)^{p}$, where $p$ is an integer. Then we can work with the following amplitude
	\begin{equation}\label{modM}
		\mathcal{M}\left(s,t\right) = \left(st\right)^{p}\mathcal{\mathcal{I}}\left(s,t\right)
	\end{equation}
	The $s$-channel discontinuity of $\mathcal{M}$ is given by 
	\begin{equation}
		\mathcal{A}^{(s)} = \left(st\right)^{p}\Biggl\{\pi \left(\frac{1}{\epsilon}-\log\left(-t\right)\right)\frac{1}{t\sqrt{s\left(s-4\right)}}\Theta\left(s-4\right) +  \frac{\pi}{s\sqrt{t\left(t-4\right)}}\log\left(1-\frac{t-\sqrt{t\left(t-4\right)}}{2}\right)\Biggr\}, \quad s>0.
	\end{equation}
	For $p>1$, we have to apply higher subtracted dispersion relation given in \eqref{HS-method1}. This leads to the following table:
    \begin{table}[H]
        \centering
        \begin{tabular}{c|c|c}
        $s$ & $t$ & $r$\\
        \hline
 $-2.92857-i 2.77273$  & $-1.84375+ i 3.62963$  & 1 \\
 $-0.435897+ i0.586207$  & $-2.27083- i0.180328$  & 1 \\
 $-2.50345+i 0.652174$  & $1.77419 +i 3.15789$  & 1 \\
 $-1.26471-i 2.36667$  & $-2.32967-i 1.9$  & 1 \\
 $-2.68571+i 1.25714$ i & $1.9959 -i 2.24324$  & 1 \\
\end{tabular}
        \caption{Results presented for the case $p=2$. The lowest order of subtraction required here is $\mathcal{O}\left(\left(s+\lambda\right)\left(t+\lambda\right)\right)$ and $\epsilon=0.01$. $\lambda$ is chosen to be $-1.5+i 10-8$. $r$ denotes the ratio of the evaluated result to the actual answer. }
    \end{table}

\section{Extended domain of convergence for string amplitudes}
\label{sec:string}
Typically, while deriving dispersion relations, we assume that the amplitude satisfies polynomial boundedness, i.e., $\displaystyle \lim_{|s| \to \infty} |\mathcal{M}(s,t)/s^N| = 0$ for some non-negative integer $N$. However, string amplitudes are special in that they exhibit Regge behaviour at high energies.
\begin{equation}
             \mathcal{M}(s,t) \sim s^{\alpha_0 + \alpha' t}\,.
\end{equation} 
Does this special behaviour buy us improved convergence properties? The answer is yes. In fact, we find that 
\begin{center}
    \textit{Amplitudes exhibiting Regge behaviour can always be expressed via the unsubtracted, parametric, crossing symmetric dispersion relation that converges everywhere, regardless of their Regge intercept.}
\end{center}
Using unsubtracted fixed-$t$ dispersion relations, we get convergence only for $\text{Re}~t < - \alpha_0/\alpha'$. However, we saw that the unsubtracted parametric CSDRs converge whenever $\displaystyle \lim_{|\sigma| \to \infty} \hat{s}_2^{\lambda}(\sigma,s_i) < - \alpha_0/\alpha'$ and since $ \displaystyle \lim_{|\sigma| \to \infty} \hat{s}_2^{\lambda}(\sigma,s_i) = -\lambda$ is independent of $s_1, s_2$, we get convergence for all values of $s_1$ and $s_2$ when $\text{Re}~\lambda > \alpha_0/\alpha'$. Quite strikingly, this is true even if the linear function in $t$ is replaced by a polynomial. \\
The extended domain of convergence becomes especially important when considering the forward limit, $t \to 0$. For example, for dilaton scattering, the Virasoro-Shapiro amplitude (given in \eqref{AmpVS}) grows as $\mathcal{M}_{VS}(s,t) \underset{s \to \infty}{\sim} s^{2+2t}$. So, the unsubtracted fixed-$t$ only converges for $t < -1$, and the twice-subtracted fixed-$t$ only for $t < 0$, both missing the forward limit. On the other hand, the unsubtracted three-channel, parametric CSDR converges for all $s$ and $t$ for $\text{Re}~\lambda > 1$. This also allows us to expand the amplitude around $s,t = 0$ and obtain dispersive representations for all Wilson coefficients. Thus, we can say 
\vspace{0.2cm}
\newline
    \textit{For amplitudes exhibiting Regge behaviour, all the Wilson coefficients are dispersive, regardless of the Regge intercept.}

    \vspace{0.2cm}
\noindent
Let us now discuss some examples of string amplitudes in more detail.

\subsection{The Veneziano amplitude} 
Applying the two-channel parametric CSDR, we get the series expansion of the Veneziano amplitude given in \eqref{VenAmpPCSDR}. We present it here again for the reader's convenience. 
\begin{equation}
\frac{\Gamma(-s)\Gamma(-t)}{\Gamma(-s-t)}=\sum_{n=0}^\infty \frac{1}{n!}\left(\frac{1}{n-s}+\frac{1}{n-t}-\frac{1}{n+\lambda}\right)\left(1-\lambda+\frac{(s+\lambda)(t+\lambda)}{n+\lambda}\right)_n\,.
\end{equation}
This expansion has previously been discussed in detail in \cite{SS, appl4}. The large $n$ behaviour of the summand is $n^{-1-\lambda}$ and therefore, the series converges for $\lambda > 0$. The fixed-$t$ dispersive representation is obtained by substituting $\lambda=-t$ and converges only for $ t < 0$. Let us consider the fixed-angle, large-energy scattering limit. Keeping the first 10 terms leads to the following plots for $s=8.8$.\\
\begin{figure}[H]
  \centering
  \begin{subfigure}[b]{0.32\textwidth}
    \includegraphics[width=\textwidth]{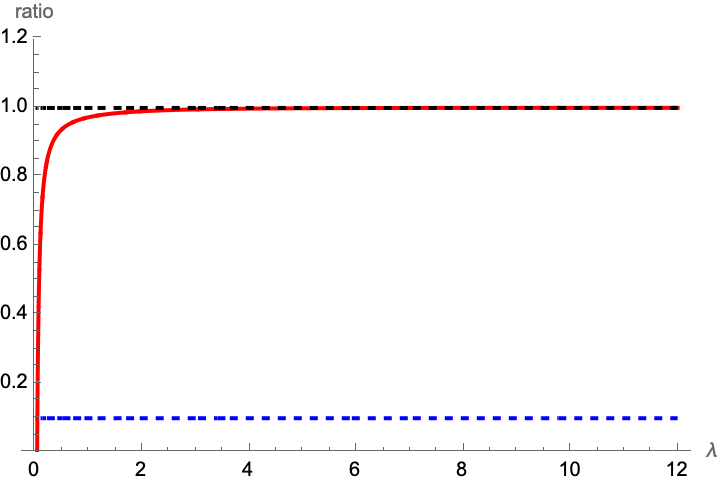}
    \caption{$z=0.99$}
  \end{subfigure}
  \begin{subfigure}[b]{0.32\textwidth}
    \includegraphics[width=\textwidth]{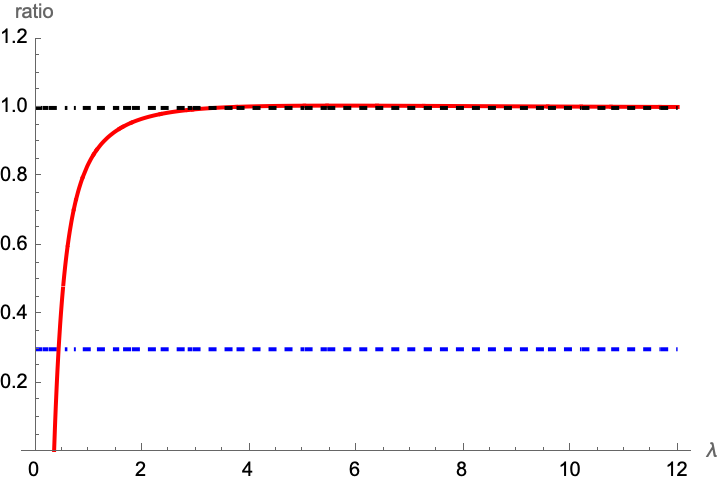}
    \caption{$z=0.9$}
  \end{subfigure}
  \begin{subfigure}[b]{0.32\textwidth}
    \includegraphics[width=\textwidth]{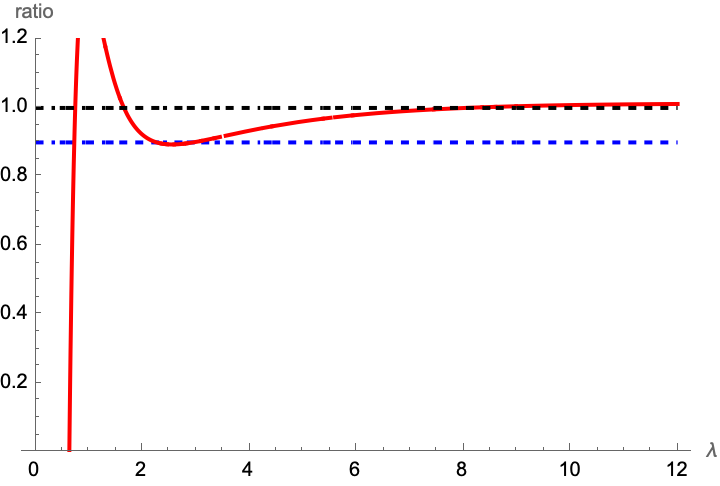}
    \caption{$z=0.5$}
  \end{subfigure}
  \caption{Ratio of the series to actual answer vs $\lambda$ in the fixed angle case. The black dashed line is 1, the blue dashed line is the fixed-$t$ series. The red line is the stringy representation.}
\end{figure}
\noindent We observe that a large number of terms are required in the fixed-$t$ series to get an accuracy similar to the parametric CSDR. For instance, while fixed-$t$ gives a $92\%$ accuracy with 500 terms, the parametric CSDR with $\lambda = 5$ gives $99\%$ accuracy with just 10 terms. 

\subsection{Virasoro-Shapiro amplitude: Dilaton scattering}

\begin{figure}[hbt]
  \centering
  \begin{subfigure}[b]{0.6\textwidth}
    \includegraphics[width=\textwidth]{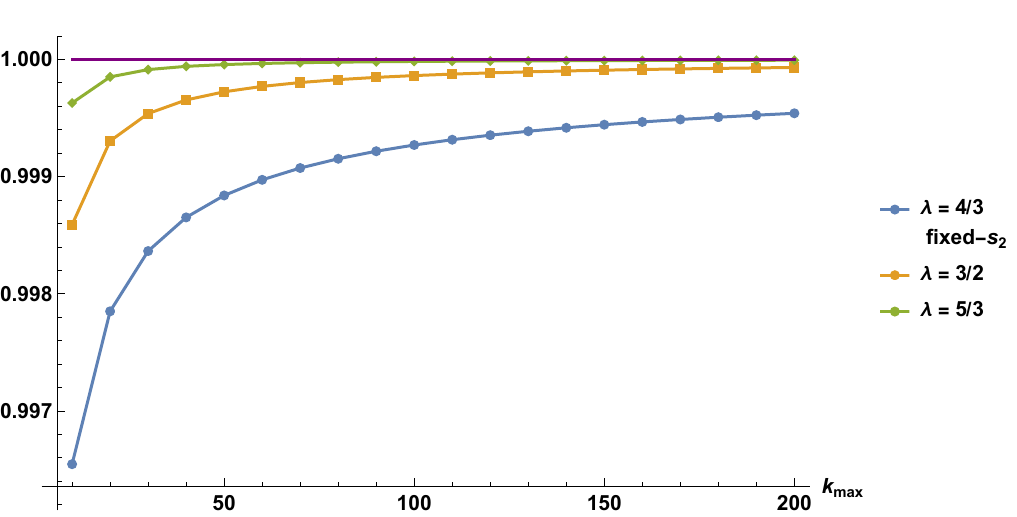}
    \caption{$s_2 < -1: s_1 = 1/2, s_2 = -4/3$}
  \end{subfigure}
  
  \vspace{0.5cm}
  
  \begin{subfigure}[b]{0.6\textwidth}
    \includegraphics[width=\textwidth]{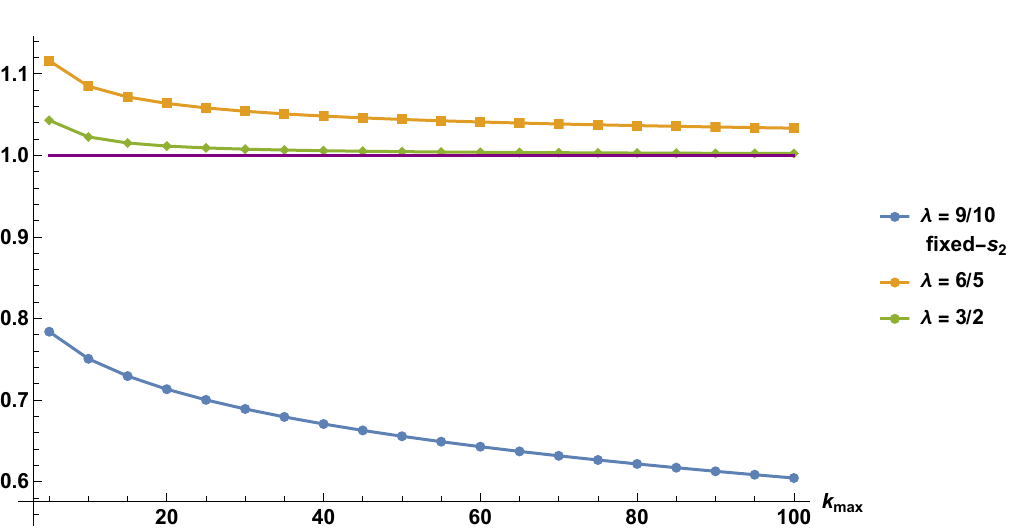}
    \caption{$s_2 > -1: s_1 = 1/2, s_2 = -9/10$}
  \end{subfigure}
 \caption{Ratio of the dispersive representation of the dilaton scattering amplitude and the exact answer for (a) $s_2 < -1$, (b) $s_2 > -1$ and various $\lambda$. We observe that the fixed-$t$ representation $(\lambda = -s_2)$ doesn't converge for case (b), while the parametric CSDR does, for $\lambda > 1$. We also see that the parametric CSDR is independent of $\lambda$, but dialing $\lambda$ improves the convergence rate.}
  \label{fig:VSDilDispPlot}
\end{figure}
\noindent The series expansion for the Virasoro-Shapiro amplitude for graviton scattering in type II string theories was discussed in \cite{appl4}. We now consider the Virasoro-Shapiro amplitude for dilaton scattering.
\begin{equation}
\label{AmpVS}
    \mathcal{M}_{VS}(s_1,s_2,s_3) =(s_1 s_2 +s_2 s_3 + s_3 s_1)^2 \frac{\Gamma(-s_1)\Gamma(-s_2)\Gamma(-s_3)}{\Gamma(1+s_1)\Gamma(1+s_2)\Gamma(1+s_3)}
\end{equation}
The $s$-channel discontinuity is given as 
\begin{equation}
   \mathcal{A}_{VS}^{(s_1)}(s_1,s_2,s_3)= -\pi\sum_{k =0}^{\infty}\frac{(-1)^{k}  }{(k!)^2 } (k s_2+s_2s_3+ ks_3)^2\frac{\Gamma (-s_2) \Gamma (-s_3)}{\Gamma (s_2+1) \Gamma (s_3+1)} \delta(k-s_1)
\end{equation}
We get the following dispersive representation by substituting this expression into \eqref{3ch_CSDisp_final1}.
\begin{equation}
\label{Dilaton_Disp}
\begin{split}
  &\mathcal{M}_{VS}(s_1,s_2,s_3) = \frac{x^2}{y}+\frac{2x}{\lambda}+ \frac{y}{\lambda^2} \\&
  -\sum_{k=1}^{\infty}\left(\frac{1}{k-s_1}+\frac{1}{k-s_2}+\frac{1}{k-s_3}- \frac{1}{k+\lambda} \right)\mathcal{A}_{VS}^{(s_1)}(k,\hat{s}_2^{(\lambda)}(k,s_i),\hat{s}_3^{(\lambda)}(k,s_i))   
\end{split}
\end{equation}
where $x = -(s_1 s_2 + s_2 s_3 + s_3 s_1)$ and $y = -s_1 s_2 s_3$ were defined previously.


\noindent The series converges for $\text{Re}~\lambda > 1$. We have separated the $ k=0$ piece as it has a simple form and gives the graviton pole term. We demonstrate the convergence of this representation and advantages over fixed-$t$ dispersion relation in Fig. \ref{fig:VSDilDispPlot}.

\subsection{Dispersive Partial Wave Expansion}
\begin{figure}[t]
    \centering
    \includegraphics[width=0.7\linewidth]{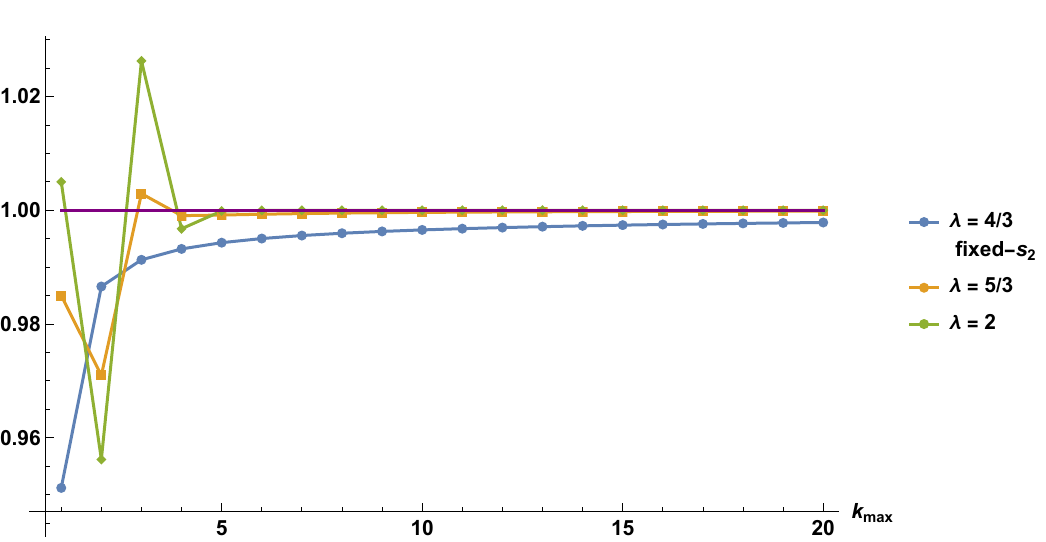}
    \caption{Ratio of the dispersive partial wave expansion of the dilaton scattering amplitude and the exact answer at $s_1 = 1/2, s_2 = -4/3$ in $d =5$ for various $\lambda$. We observe that the expansion is $\lambda$-independent, but by dialling $\lambda$, we can improve the convergence rate, and in particular, get faster convergence compared to the fixed-$t$ dispersion relation $(\lambda = -s_2)$.}
    \label{fig:VSDialPWEPlot}
\end{figure}

The partial wave expansion of a 2-2 scattering amplitude of identical scalars of mass $m$ in $d$ dimensions is given as
\begin{equation}
\label{PWE}
    \mathcal{M}(s,t) = \sum_{\ell =0}^{\infty}  f_{\ell}(s) \mathcal{C}^{((d-3)/2)}_{\ell}(z) , \quad z = 1+ \frac{2t}{s - 4m^2} \,.
\end{equation}
The $f_{\ell}(s)$ are the spin $\ell$ partial wave coefficients and $\mathcal{C}^{(\frac{d-3}{2})}_{\ell}(z)$ are the $d$-dimensional Gegenbauer polynomials. 
\begin{equation}
\label{ch2_gegdef}
  \mathcal{C}^{(\alpha)}_{\ell}(z)  =\frac{\Gamma(\ell + 2\alpha)}{\Gamma(2 \alpha)~ \ell!} ~ {}_2F_1\left(-\ell,\, \ell + 2\alpha;\, \alpha + \frac{1}{2};\, \frac{1 - z}{2} \right),
\end{equation}
We can extract the $f_{\ell}(s)$ using orthogonality properties of Gegenbauer polynomials. 
We can also expand the discontinuity of the amplitude in terms of partial waves. Via the dispersion relation, this lets us express the full amplitude only in terms of the imaginary part of the partial waves. Let us see how this works for the Dilaton scattering amplitude. The  discontinuity can be expanded as 
\begin{equation}
    \mathcal{A}_{VS}^{(s_1)}(s_1,s_2,s_3)= \pi\sum_{k=0}^{\infty}\sum_{\ell =0}^{2k+2} \text{Im} f_{\ell}(k) \mathcal{C}^{((d-3)/2)}_{\ell}\left(1+ \frac{2 s_2}{k}\right)\delta(k-s_1)
\end{equation}
Only even spins contribute, as we have identical scalar particles scattering. Plugging this into the dispersive representation \eqref{Dilaton_Disp}, we get the following partial wave expansion
\begin{equation}
\label{Dilaton_Disp_PWE}
\begin{split}
  &\mathcal{M}_{VS}(s_1,s_2,s_3) = \frac{x^2}{y}+\frac{2x}{\lambda}+ \frac{y}{\lambda^2} \\&
  -\sum_{k=1}^{\infty}\left(\frac{1}{k-s_1}+\frac{1}{k-s_2}+\frac{1}{k-s_3}- \frac{1}{k+\lambda} \right)  \sum_{\ell =0}^{2k+2} \text{Im} f_{\ell}(k) \mathcal{C}^{((d-3)/2)}_{\ell}\left(1+ \frac{2 \hat{s}_2^{(\lambda)}(k,s_i)}{k}\right)    
\end{split}
\end{equation}
We demonstrate the convergence of this expansion in Fig. \ref{fig:VSDialPWEPlot}.

\subsection{Bounds on weakly-coupled gravitational EFTs}
\label{g2g3bounds}
Consider the 2-2 scattering of identical massless scalars coupled to gravity. We assume that the low-energy EFT is weakly interacting, but remain agnostic about the physics above the EFT cut-off $\Lambda$, except that unitarity, analyticity, and crossing symmetry continue to hold. Below the cut-off, the EFT is parametrised by Wilson coefficients that appear in the low-energy expansion of the amplitude around $x, y = 0$.
\begin{equation}
    \mathcal{M}(s,t) = \frac{x^2}{y} + \sum_{p,q = 0}^{\infty} W_{p,q} x^p y^q \,.
\end{equation}
$W_{p,q}$ are the Wilson coefficients. Let us assume that the amplitude satisfies $\displaystyle \lim_{|s| \to \infty}\left|\mathcal{M}(s,t)/s^2\right| = 0, \text{ for } t<0 $. Then the amplitude can be represented using the twice-subtracted parametric CSDR given in \eqref{TwiceSubCSDR}, where the $s$-channel discontinuity starts above the EFT cut-off $\Lambda$. The coefficient $W_{0,0}$ can be taken as the subtraction constant in this case and is therefore non-dispersive. To obtain dispersive representations of other Wilson coefficients, we need to Taylor-expand the parametric CSDR around $x,y =0$. For non-gravitational EFTs, this expansion is possible to perform using the fixed-$t$ dispersion relation. Bounds on the Wilson coefficients are then obtained by plugging in the partial wave expansion in their dispersive representations and imposing unitarity above the EFT cut-off. However, it is well known that for gravitational EFTs, this way of deriving bounds fails due to the presence of the graviton pole, which prevents the expansion of the fixed-$t$ dispersion relation in the forward limit, $t \to 0$. In \cite{Swamp}, a method to bypass this problem was proposed by deriving bounds in the limit of a small impact parameter, rather than the forward limit. This problem can be avoided if we use the twice-subtracted, parametric CSDR. Given the assumed high-energy behaviour, we expect convergence for all $s$ and $t$ for $\lambda > 0$, and therefore, can expand the parametric CSDR around the forward limit. The divergence from the graviton pole is regulated by choosing some non-zero value for $\lambda$. 

\begin{figure}[H]
    \centering
    \includegraphics[width=0.8\linewidth]{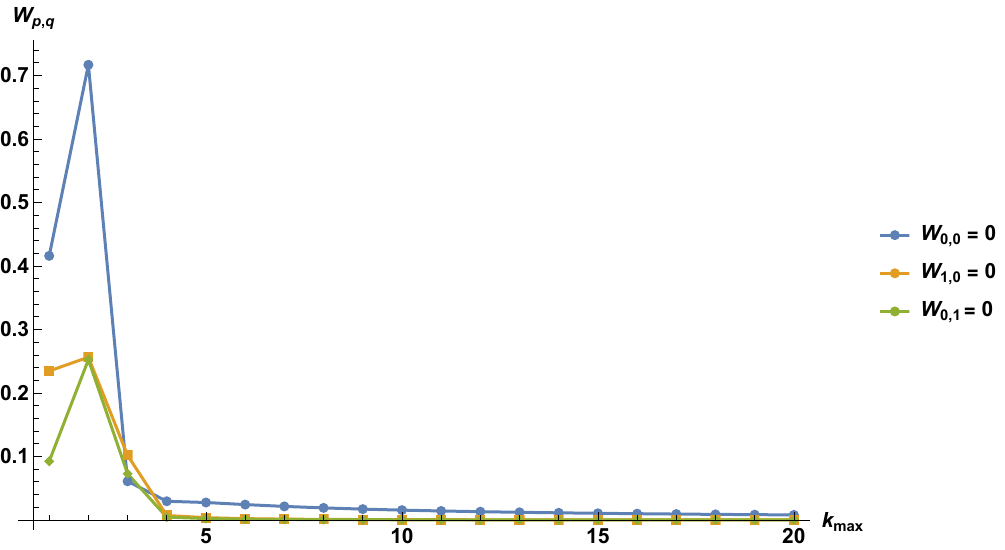}
    \caption{Dispersive partial wave expansions of the Wilson coefficients $W_{0,0},W_{1,0},W_{0,1}$ for the dilaton scattering amplitude for $\lambda = 3/2$ in $d =5$. }
    \label{fig:WPQPWEPlot}
\end{figure}

We shall now demonstrate this procedure by studying the space of dilaton-like scattering amplitudes. We fix the spectrum to be the same as for the dilaton amplitude in \eqref{Dilaton_Disp} and assume the same high-energy behaviour. $k=1$ corresponds to the first massive exchange. As mentioned before, due to Regge behaviour at large $s$, the unsubtracted CSDR suffices and all the Wilson coefficients, including $W_{0,0}$, are dispersive. Using \eqref{Dilaton_Disp_PWE}, we get the following dispersive partial wave expansions for the first few Wilson coefficients, which converge for $\lambda > 1$.
\begin{equation}
\label{WpqPWE}
\begin{split}
 W_{0,0} &= -\sum_{k =1}^{\infty}\sum_{\ell =0}^{2k+2} \text{Im} f_{\ell}(k)\mathcal{C}_{\ell}^{\left(\frac{d-3}{2}\right)}\left( \sqrt{\frac{k-3\lambda}{k+\lambda}}\right)\left(\frac{3}{k}- \frac{1}{k+\lambda}\right) \,,\\   
 W_{1,0} &= \frac{2}{\lambda}-\sum_{k =1}^{\infty}\sum_{\ell =0}^{2k+2} \text{Im} f_{\ell}(k)\frac{2}{k^3}\left[C_{\ell}^{\left(\frac{d-3}{2}\right)}\left(\frac{\sqrt{k-3 \lambda }}{\sqrt{k+\lambda }}\right)+\frac{(d-3) \lambda  (2 k+3 \lambda ) }{\sqrt{k-3 \lambda } (k+\lambda )^{3/2}}C_{\ell-1}^{\left(\frac{d-1}{2}\right)}\left(\frac{\sqrt{k-3 \lambda }}{\sqrt{k+\lambda }}\right)\right]\,,\\
  W_{0,1} &= \frac{1}{\lambda^2}+\sum_{k =1}^{\infty}\sum_{\ell =0}^{2k+2} \text{Im} f_{\ell}(k)\frac{1}{k^4}\left[3 C_{\ell}^{\left(\frac{d-3}{2}\right)}\left(\frac{\sqrt{k-3 \lambda }}{\sqrt{k+\lambda }}\right)-\frac{2 (d-3) k (2 k+3 \lambda )}{\sqrt{k-3 \lambda } (k+\lambda )^{3/2}} C_{\ell-1}^{\left(\frac{d-1}{2}\right)}\left(\frac{\sqrt{k-3 \lambda }}{\sqrt{k+\lambda }}\right)\right]\,.
\end{split}
    \end{equation}
All three of these Wilson coefficients are zero for dilaton scattering. We demonstrate convergence of their partial wave expansions in Fig. \ref{fig:WPQPWEPlot}. 

To derive bounds, we need to impose:
\begin{itemize}
    \item \textit{Tree-level Unitarity}:  $\text{Im} f_{\ell}(k) \ge 0$ for all $k \ge 1$ and even spins with $0 \le \ell \le 2k+2$.
    \item \textit{$\lambda$-independence}: These are constraints that come from demanding that the dispersive partial wave expansions be independent of the parameter $\lambda$ in the convergence domain $\text{Re} (\lambda) > 1$ (this condition is equivalent to the null constraints in the fixed-$t$ approach). 
\end{itemize} 
We shall focus on obtaining bounds in the $W_{1,0}$-$W_{0,1}$ plane by using a dual optimisation setup. Let us define
\begin{equation}
    \tilde{W}_{1,0}(\lambda) = W_{1,0} - \frac{2}{\lambda} \quad and \quad \tilde{W}_{0,1}(\lambda) = W_{0,1} - \frac{1}{\lambda^2}
\end{equation}
Then from \eqref{WpqPWE}, these are given as
\begin{equation}
\begin{split}
    \tilde{W}_{1,0}(\lambda)  = \sum_{k =1}^{k_{max}}\sum_{\ell =0}^{2k+2} \text{Im} f_{\ell}(k) \tilde{w}_{1,0}(\ell,k,\lambda), \quad   \tilde{W}_{0,1}(\lambda)  = \sum_{k =1}^{k_{max}}\sum_{\ell =0}^{2k+2} \text{Im} f_{\ell}(k) \tilde{w}_{0,1}(\ell,k,\lambda),
\end{split}
\end{equation}
where we have truncated the $k$-sum to some large enough $k_{max}$. Now, suppose we try to impose, for some $\alpha, \beta \in \mathbb{R}$,
\begin{equation}
   \alpha ~ \tilde{w}_{0,1}(\ell,k,\lambda) + \beta ~ \tilde{w}_{1,0}(\ell,k,\lambda) \ge 0, \quad \forall ~ k,l.
\end{equation}
It can be shown that there is no choice of $\lambda$ such that $\text{Re}(\lambda) \ge 1$ for which this holds. This follows from the fact that in \eqref{WpqPWE}, the Gegenbauer polynomials are only positive when their argument is greater than 1. Therefore, to get bounds, just unitarity is not sufficient; we need to supplement the $\lambda$-independence constraints as well. We do this as follows. We pick a grid of $\lambda$ values satisfying $\lambda \in (1, \lambda_{max}]$ - let us refer to the grid as $\lambda_{grid}$. Then we impose positivity, which leads to \footnote{This procedure is similar to the "smearing" procedure in \cite{Swamp}.}
\begin{equation}
\begin{split}
     & \sum_{\lambda_i \in \lambda_{grid}}  \alpha(\lambda_i) ~ \tilde{w}_{0,1}(\ell,k,\lambda_i) + \beta(\lambda_i) ~ \tilde{w}_{1,0}(\ell,k,\lambda_i) \ge 0, \quad \forall ~ k,l. \\
     & \implies \sum_{\lambda_i \in \lambda_{grid}} \alpha(\lambda_i) ~ \tilde{W}_{0,1}(\lambda_i) + \beta(\lambda_i) ~ \tilde{W}_{1,0}(\lambda_i) \ge 0, \quad \text{(Unitarity)} \\
     &\implies \sum_{\lambda_i \in \lambda_{grid}} \alpha(\lambda_i) \left(W_{0,1} -\frac{1}{\lambda_i^2} \right) + \beta(\lambda_i) \left(W_{1,0} -\frac{2}{\lambda_i} \right)  \ge 0, \quad \text{($\lambda$-independence)} \\
     &\implies W_{0,1}  \left(\sum_{\lambda_i \in \lambda_{grid}} \alpha(\lambda_i) \right) \ge   \sum_{\lambda_i \in \lambda_{grid}} \alpha(\lambda_i) \frac{1}{\lambda_i^2}- \beta(\lambda_i) \left(W_{1,0} -\frac{2}{\lambda_i}  \right) 
\end{split} 
\end{equation}
Then, if we demand $\displaystyle\sum_{\lambda_i \in \lambda_{grid}} \alpha(\lambda_i)  = 1$, we can find a lower bound on $W_{0,1}$ for any choice of $W_{1,0}$ by maximizing the RHS. 
\begin{equation}
    \text{min}(W_{1,0}) \ge \text{max}\left( \sum_{\lambda_i \in \lambda_{grid}} \alpha(\lambda_i) \frac{1}{\lambda_i^2}- \beta(\lambda_i) \left(W_{1,0} -\frac{2}{\lambda_i}  \right)  \right)
\end{equation}
A similar procedure also leads to an upper bound on $W_{0,1}$ for any choice of $W_{1,0}$. We impose these constraints using the semi-definite optimisation package SDPB \cite{SDPB}. Details of the numerics and SDPB implementation are given in the Appendix. We present our results for $d = 5$ in Fig. \ref{fig:g2g3plot}.

\begin{figure}[H]
    \centering
    \includegraphics[width=0.8\linewidth]{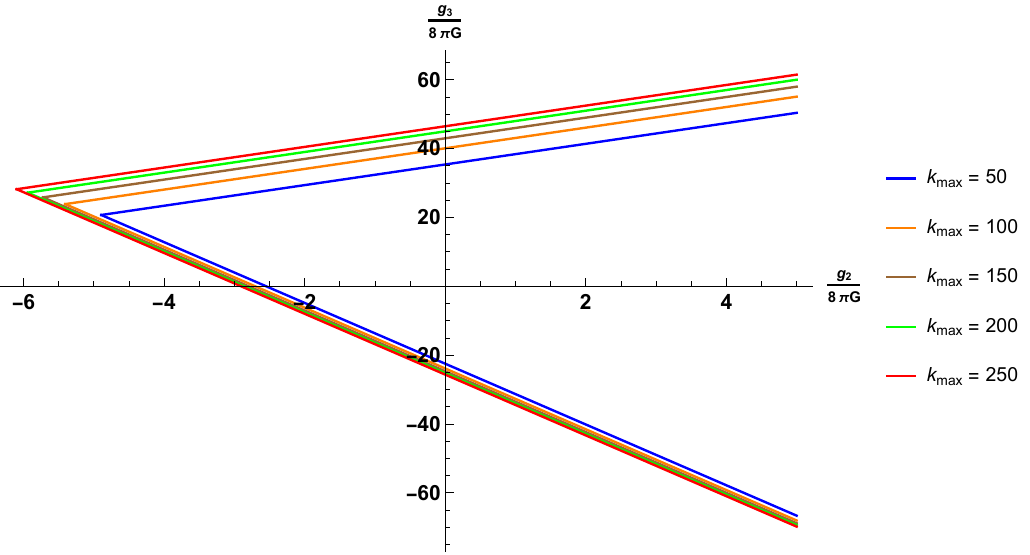}
    \caption{Bounds with gravity for dilaton-like amplitudes in $d = 5$. The region to the right of the curves is allowed. As can be seen, the bounds converge as $k_{max}$ is increased. We define $W_{1,0} = - \dfrac{2g_2}{8 \pi G}$ and $W_{0,1} = \dfrac{g_3}{8\pi G}$ to match the conventions in \cite{Swamp} and aid comparison. Note that the bounds here are stronger compared to \cite{Swamp} because we only a consider a restricted subspace of meromorphic amplitudes.}
    \label{fig:g2g3plot}
\end{figure}

\section{Towards an $n$-particle dispersion relation}
\label{sec:npoint}
In this section, we will exhibit a totally symmetric $n$-variable dispersion relation and outline a strategy to use it for $n$-particle dispersion relations with full crossing symmetry. In the $n$-particle case, we need to be careful in distinguishing the totally symmetric functions from the crossing symmetric functions---we will clarify this below. 
Consider the multivariable generalisation of the Virasoro-Shapiro amplitude:
\begin{equation}
{\mathcal M}_{VS}(\{s_i\})=\prod_{i=1}^n \frac{\Gamma(-s_i)}{\Gamma(1+s_i)}\,.    
\end{equation}
This has at $s_i=n_i$ for integer $n_i\geq 0$. Alternatively, as another example where the variables are coupled, consider the multivariable generalisation of the Euler-Beta function of the form
\begin{equation}
    {\mathcal M}_V(\{s_i\})=\prod_{i=1}^n \frac{\Gamma(-s_i)}{\Gamma(-\sum_{i=1}^n s_i)}\,.
\end{equation}
Neither of these amplitudes actually makes an appearance in any known physical theory to the best of our knowledge. However, they will enable us to chase what we are after; namely, a totally symmetric $n$-variable dispersion relation. 

 Our considerations in the previous sections lead us to conjecture the following dispersive formula:
\begin{equation}
{\mathcal M}(\{ s_i \})=\int d\sigma \mathcal{H}^{(\lambda)}(\sigma,\{s_i\})\mathcal{A}^{(s_1)}(\sigma, \{r_i\})\,,    
\end{equation}
where $\mathcal{A}^{(s_1)}$ is the discontinuity in the $s_1$ variable. 
\begin{equation}
    \mathcal{H}^{(\lambda)}(\sigma,\{s_i\})=\frac{1}{\sigma+\lambda}+\sum_{i=1}^n \frac{1}{s_i-\sigma} \,,
\end{equation}
and $r_i$ are the $n-1$ nontrivial (ignoring the $\tau=\sigma$) roots of $\tau$, depending on $\sigma$ and totally symmetric combinations of $s_i$, of
\begin{equation}
    \frac{\prod_{i=1}^n (s_i-\sigma)}{\sigma+\lambda}=\frac{\prod_{i=1}^n (s_i-\tau)}{\tau+\lambda}\,.
\end{equation}
This representation makes the single poles in each variable manifest in a crossing symmetric manner. We have explicitly checked that this representation works for $\mathcal{M}_{V/VS}(s_i)$ for up to $n=6$. Explicit formulas for $r_i$ can be found using Mathematica for up to\footnote{The $n=5$ case involves finding a solution to $(\tau-\sigma)p_4(\tau)=0$, where $p_4$ is a 4-th order polynomial. Hence, explicit (ugly) solutions exist.} $n=5$, after which the formulas are implicit. However, this does not prevent us from going to any $n$. In principle, one can also add more $1/(\sigma+\lambda_i)$ terms to the kernel, which will enable an extrapolation to the case where all except one variable are held fixed. We will not consider this case. 

\subsection{5 particle crossing symmetric case: Take one}

Let us consider the 5-particle case. Here, we will have 10 variable analogues of the Mandelstam invariants given by $(p_i+p_j)^2=s_{ij}$ with $i,j=1,2\cdots,5$. Kinematic constraints reduce the number of independent variables to 5. For instance, in \cite{bardakci}, these are chosen to be $s_{12},s_{23},s_{34},s_{45}, s_{51}$. 
Let us be pedantic at this point and consider the massless $\phi^3$ theory as an example. Here using momentum conservation, it is easy to show that $s_{13}=s_{45}-s_{12}-s_{23}, s_{14}=s_{23}-s_{15}-s_{45}, s_{24}=s_{15}-s_{23}-s_{34}, s_{25}=s_{34}-s_{12}-s_{15}, s_{35}=s_{12}-s_{34}-s_{45}\,.$ Crossing symmetry implies that in terms of the momenta labels, the amplitude is permutation symmetric. However, this does not imply total symmetry in terms of either the 10 variables or the 5 independent variables! To see this clearly, consider the tree-level amplitude, which is given by
\begin{eqnarray}
{\mathcal M}(s_{ij})=\frac{1}{s_{12}}\left(\frac{1}{s_{34}+s_{35}+s_{45}}\right)+\cdots
\end{eqnarray}
where the $\cdots$ are obtained by permuting the momenta labels. Crucially, we do not have terms like $1/s_{12}s_{13}$, and hence the amplitude is not totally symmetric. 
The algorithm to utilize the totally symmetric dispersion relations would be to consider an extension of an old idea from 1969 due to Roskies \cite{roskies}, which was also used recently in \cite{appl1}. In \cite{roskies} it was shown how to expand any $F(s,t,u)$ function (with no symmetry in $s,t,u$) that is analytic in a domain which is invariant under permutations of $s,t,u$ in terms\footnote{The domain is invariant under permutations of $s,t,u$, not the function $F(s,t,u)$!} of totally symmetric functions using the representations of $S_3$. In \cite{appl1}, this idea was used to write crossing symmetric dispersion relations (in 2-2 scattering of identical scalars, totally symmetric and crossing symmetric are used interchangeably) for amplitudes of particles carrying spin. 

\begin{figure}[H]
    \centering
    \includegraphics[width= 0.6\linewidth]{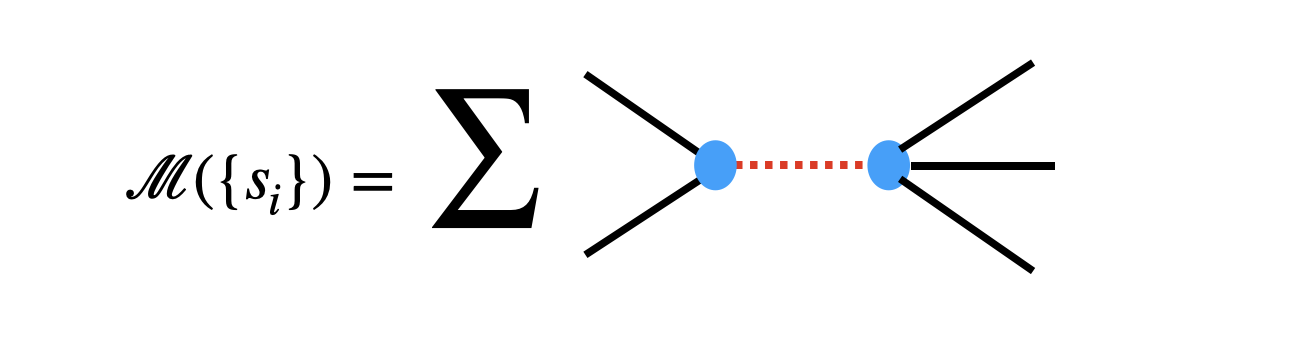}
    \caption{Take 1: The 5-particle case shown schematically. }
    \label{fig:5disp}
\end{figure}

\noindent We outline the next steps for future investigations. We will need to expand ${\mathcal M}(s_{ij})$, which depends on the 10 variables, in terms of a basis is in terms of totally symmetric functions as in \cite{roskies}. Then the $n$-variable dispersion relations discussed above can be applied to these totally symmetric functions. Finally, the kinematic constraints can be imposed at the end. At this stage, we have a decomposition that is shown schematically in Fig. \ref {fig:5disp}. 

 Now note that since $r_i$'s are functions of $s_i$ involving roots, $\mathcal{A}^{(s_1)}$ in this case may hide further singularities which come from $s_i, i>1$ in ${\mathcal M}(\{s_i\})$. To take the next step in this program, one will have to find formulas for expanding $\mathcal{A}^{(s_1)}$. This may enable one to find a connection with \cite{nima}. We hope to come back to this in the future.

\subsection{5 particle case: Take two}
The approach above, while implementable, is probably not the most efficient one. Here, we will try to proceed as closely as possible to the 2-2 case for the Veneziano amplitude to come up with the necessary ingredients to write down a similar dispersive representation for the 5-particle analogue of the Veneziano amplitude (with the appropriate fall-off). First, we need the analogue of the kernel, which has only cyclic symmetry in the independent variables $s_{12},s_{23},s_{34},s_{45},s_{51}$. A moment's thought suggests that we need to introduce at least two integration variables. The following appears to be the best candidate:
\begin{eqnarray}
{\mathcal{H}^{\lambda_i}(\sigma,\tau;\{s_{ij}\}})&=&\frac{1}{(s_{12}-\sigma)(s_{34}-\tau)}+   \frac{1}{(s_{23}-\sigma)(s_{45}-\tau)}+\frac{1}{(s_{34}-\sigma)(s_{51}-\tau)}\nonumber\\&+&\frac{1}{(s_{45}-\sigma)(s_{12}-\tau)}+\frac{1}{(s_{51}-\sigma)(s_{23}-\tau)}-\Biggl\{\frac{2}{\left(\lambda+\sigma\right)\left(\lambda+\tau\right)} \nonumber\\
&& \frac{1}{\left(\lambda+\sigma\right)\left(\mu_{1}+\tau\right)} + \frac{1}{\left(\mu_{2}+\sigma\right)\left(\lambda+\tau\right)}\Biggr\} \,.
\end{eqnarray}
The reason we have introduced the last term is to recover the fixed-three-variable dispersion relation. For example, if $s_{23}$, $s_{45}$ and $s_{51}$ are fixed to $-\lambda$ then the kernel reduces to $\mathcal{H}=\frac{1}{(s_{12}-\sigma)(s_{34}-\tau)}$, provided in that case $\mu_{1}\rightarrow s_{12}$ and $\mu_{2}\rightarrow s_{34}$. Here $\mu_{1}$ and $\mu_{2}$ are not independent variables and have functional dependencies on $\lambda$.

The next step is to write a 2-variable Cauchy integral of this kernel times the amplitude, which now takes the form
\begin{equation}
    {\mathcal M}(\sigma,s'_{23}(\sigma,\tau),\tau,s'_{45}(\sigma,\tau),s'_{51}(\sigma,\tau))\,,
\end{equation}
where the functional dependence of $s'_{23}, s'_{45}, s'_{51}$ is also on all the independent $s_{ij}$'s which we have suppressed for notational simplicity. The condition from the Cauchy contour picking up the $\sigma, \tau$ poles from the kernel gives us:
\begin{eqnarray}\label{prop1}
    s'_{23}(s_{12},s_{34})&=&s_{23}\,,\quad s'_{45}(s_{12},s_{34})=s_{45}\,,\quad s'_{51}(s_{12},s_{34})=s_{51}\,,\\
    s'_{23}(s_{23},s_{45})&=&s_{34}\,,\quad s'_{45}(s_{23},s_{45})=s_{51}\,,\quad s'_{51}(s_{23},s_{45})=s_{12}\,,\\
     s'_{23}(s_{34},s_{51})&=&s_{45}\,,\quad s'_{45}(s_{34},s_{51})=s_{12}\,,\quad s'_{51}(s_{34},s_{51})=s_{23}\,,\\
     s'_{23}(s_{45},s_{12})&=&s_{51}\,,\quad s'_{45}(s_{45},s_{12})=s_{23}\,,\quad s'_{51}(s_{45},s_{12})=s_{34}\,,\\
     s'_{23}(s_{51},s_{23})&=&s_{12}\,,\quad s'_{45}(s_{51},s_{23})=s_{34}\,,\quad s'_{51}(s_{23},s_{23})=s_{45}\,.
     \end{eqnarray}
To justify the existence of equations whose solutions are the above ones, we give an example of a set of equations below: 
\begin{eqnarray*}
	s'_{23}s'_{45}s'_{51}  + \sigma s'_{23}s'_{45} + \tau s'_{23}s'_{51} + \sigma \tau \left(s'_{45}+s'_{51}\right) = s_{23}s_{45}s_{51} + s_{12}s_{34}s_{51} + s_{12}s_{23}s_{45} + s_{23} s_{34}s_{51} + s_{12}s_{34}s_{45},
\end{eqnarray*}    
\begin{eqnarray*}
	\sigma \tau \left[s'_{23}\left(s'_{45}+s'_{51}\right)+ s'_{45}s'_{51}\right]  + \left(\sigma+\tau\right)s'_{23}s'_{45}s'_{51}  &=& s_{12}s_{23}s_{34}s_{45}+ s_{23}s_{34}s_{45}s_{51} \nonumber\\
    && + s_{12}s_{34}s_{45}s_{51} + s_{12}s_{23}s_{45}s_{51} + s_{12}s_{23}s_{34}s_{51},
\end{eqnarray*}
\begin{eqnarray}
	\sigma \tau s'_{23}s'_{45}s'_{51} = s_{12}s_{23}s_{34}s_{45}s_{51}.
\end{eqnarray}
It can be verified that one of the five solutions satisfies the required property as stated in \eqref{prop1}.
\begin{figure}[H]
    \centering
    \includegraphics[width= 0.6\linewidth]{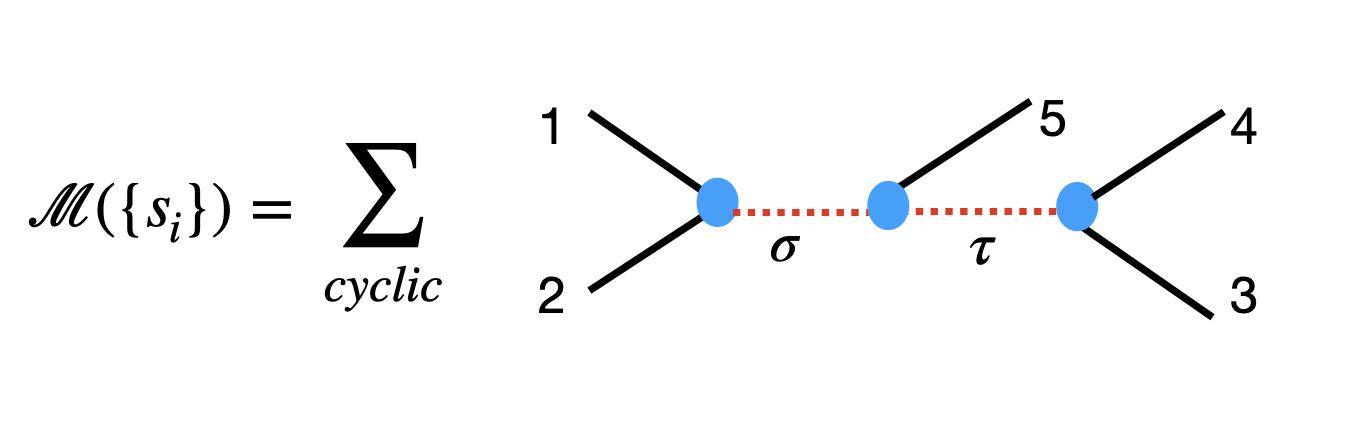}
    \caption{Take 2: The 5-particle cyclic case shown schematically. }
    \label{fig:5disp2}
\end{figure}

We assume that the amplitude $\mathcal{M}(s_{12},s_{23},s_{34},s_{45},s_{51})$ has simultaneous discontinuities in the following pairs: $(s_{12},s_{34}), (s_{12},s_{45}), (s_{23},s_{45}),(s_{23},s_{51}), (s_{34},s_{51})$. In the dual resonance models, $s_{12},s_{23},s_{51}$ are dual variables and the analog of the fixed-$t$ representation is the fixed-$s_{23},s_{51}$ representation. The absorptive part of the amplitude, using cyclic symmetry, satisfies:
\begin{eqnarray}
    &&\mathcal{A}^{(s_{12},s_{34})}(\sigma,s_{23},\tau,s_{45},s_{51})=\mathcal{A}^{(s_{23},s_{45})}(s_{51},\sigma,s_{23},\tau,s_{45})=\mathcal{A}^{(s_{34},s_{51})}(s_{45},s_{51},\sigma,s_{23},\tau)\nonumber\\&=&\mathcal{A}^{(s_{45},s_{12})}(\tau,s_{45},s_{51},\sigma,s_{23})=\mathcal{A}^{(s_{51},s_{23})}(s_{23},\tau,s_{45},s_{51},\sigma)\,.
\end{eqnarray}
The next steps, analogous to the 2-2 scattering case, are:
\begin{enumerate}
\item Take the absorptive parts corresponding to the pairwise discontinuities indicated above. For instance, consider the contribution from $(s_{45},s_{12})$ discontinuities. This will give a term
\begin{equation}\label{5part2}
    \int d\sigma d\tau \mathcal{H}^{\lambda_i}(\sigma,\tau;\{s_{ij}\})\mathcal{A}^{(s_{12},s_{34})}(s_{45}',s_{51}',\sigma,s_{23}',\tau)\,.
\end{equation}
The primes indicate that these are functions of $(\sigma,\tau, \{s_{ij}\})$. 
Here, we have used the cyclic relations between the absorptive parts to obtain the final form. We want this to be the same as what comes from the $(s_{12},s_{34})$ discontinuities:
\begin{equation}\label{5part1}
    \int d\sigma d\tau \mathcal{H}^{\lambda_i}(\sigma,\tau;\{s_{ij}\})\mathcal{A}^{(s_{12},s_{34})}(\sigma,s_{23}',\tau,s_{45}',s_{51}')\,.
\end{equation}
\item We do a change of variables $(\sigma,\tau)\rightarrow (\hat\sigma,\hat \tau)$ in eq.(\ref{5part2}) and relabel $(\sigma,\tau)\rightarrow (\hat\sigma,\hat \tau)$ in eq.(\ref{5part1}) and then equate the two integrands. This gives the analogue of eq.(\ref{conds}). Repeat for the four other pairs of discontinuities $(s_{23},s_{45}), (s_{34},s_{51}),(s_{51},s_{23})$. 
\item Solve the resulting conditions! 
\end{enumerate}
We leave the implementation and investigations of this proposal for future work. The 5-particle string amplitude will be an ideal first test-bed for this implementation. 
\section{Discussion}
\label{sec:discussion}
We conclude with some suggestions for interesting future directions.
\begin{itemize}
\item Obviously, the most important open problem is to consider the $n$-particle dispersion relation and come up with a version that enables us to find efficient series representations of the Koba-Nielsen amplitudes, which converge everywhere except at the poles. This will be in the spirit of what is expected from string field theory \cite{sen}.

\item We emphasise that simplicity was an essential criterion to choose the form of the kernel in the stringy dispersion relation. The kernel, in turn, was instrumental in fixing the ``other variables" in the absorptive part of the amplitude, which entered the dispersion relation. We pointed out that the kernel can be suitably modified while retaining the feature that it could extrapolate between the fixed-$t$ and fixed-$s$ dispersion relations. These will lead to more complicated forms of the dispersive representation. Although we did not examine such forms, it may well be that convergence is controlled by different ``slices" of the $s-t$ plane. Finding the ``geodesic" in this language may be an interesting exercise to carry out. Some partial results exist in the literature, dating back to the work of Auberson and Epele \cite{auberson}. We are aware of an ongoing attempt to further research in this direction \cite{zahedetal}.

\item In Andre Martin's derivation \cite{martin} of the enlarged analyticity domain for 2-2 scattering (pions), fixed-$t$ dispersion relations played a crucial role. An obvious question is whether the stringy dispersion relation can enlarge the domain further. Roy and Wanders had briefly examined this question in a largely ignored work in the 1970s \cite{roywanders}. It may be worthwhile to re-examine this question with the stringy dispersion relation.

\item It will be interesting to redo the pion bootstrap \cite{pion1,pion2,pion3} and the nonlinear string bootstrap \cite{string1} using the ideas in this paper.

\item Finally, as explained in section \ref{g2g3bounds}, the parametric dispersion relation allows us to derive bounds on Wilson coefficients in weakly-coupled gravitational EFTs without the explicit need to go to impact parameter space. It will be interesting to implement this idea and analyse EFT bounds in gravitational scatterings involving loop effects and infrared divergences \cite{Swamp, Caron-Huot:2022ugt, BulkLocality, appl7, deRham:2022gfe, Beadle:2025cdx}.

\end{itemize}

\section*{Acknowledgments} We thank B. Ananthanarayan, Rajesh Gopakumar, Andrea Guerrieri, Mehmet Gumus, Apratim Kaviraj, Alok Laddha, Arkajyoti Manna, Joan Miro, Prashanth Raman, Ashoke Sen and Ahmadullah Zahed for useful discussions on related topics over the last several years. We also thank Ahmadullah Zahed and Prashanth Raman for initial discussions in 2023 on higher subtracted CSDRs. We thank the organizers of QCD meets Gravity 2024, where the main idea behind this work was conceived. AS acknowledges support from a Quantum Horizons Alberta chair professorship. APS is supported by the DST INSPIRE Faculty Fellowship (IFA22-PH 282).

\appendix

\section{Parametric dispersion relation}
Our goal is to obtain the parametric dispersion relation from first principles, starting from the complex plane. This analysis uses the techniques developed in \cite{old1, song}. We choose the following parametrisation, 
\begin{equation}\label{mod-parameter}
	s_{k} = \frac{4\hat{a}ze^{i\pi k}}{\left(z+e^{i\pi k}\right)^{2}} - \lambda, \quad k=1,2.
\end{equation}
It can be checked that 
\begin{equation}\label{a-mod}
	\hat{a} = \frac{\left(s_{1}+\lambda\right)\left(s_{2}+\lambda\right)}{s_{1}+s_{2}+2\lambda},
\end{equation}
using which we will get
\begin{equation}
	s_{2} = \frac{\hat{a}\left(s_{1}+\lambda\right)}{s_{1}+\lambda-\hat{a}}-\lambda .
\end{equation}
We can invert \eqref{mod-parameter} to obtain $z$ in terms of $s_{1}$, $\lambda$ and $\hat{a}$,
\begin{equation}
	z=\frac{s_{1}+\lambda-2\hat{a}\pm 2i\sqrt{\hat{a}\left(s_{1}+\lambda-\hat{a}\right)}}{s_{1}+\lambda}.
\end{equation}
This equation implies that the singularities of the amplitude are mapped to the unit circle in the complex plane. We restrict the $s_{1}$ channel singularities to be on the upper half plane. Then we can denote the $s_{1}$ channel discontinuity of the amplitude to lie along 
\begin{equation}
	z' = e^{i\phi}, \qquad \cos\phi=\frac{\sigma+\lambda-2\hat{a}}{\sigma+\lambda}, \qquad \sin\phi= \frac{2\sqrt{\hat{a}\left(\sigma+\lambda-\hat{a}\right)}}{\sigma+\lambda}, \qquad 0\le\phi\le\pi, \quad 0<\sigma<\infty .
\end{equation}
Let us assume that the amplitude is symmetric in $s_{1}+\lambda$ and $s_{2}+\lambda$. Starting from the contour integral,
\begin{equation}
	\frac{1}{2\pi i}\oint_{|z'|>1}\frac{\mathrm{d}z'}{z'-z}\frac{z^{'2}-1}{z^{'2}}\widetilde{\mathcal{M}}\left(z',\hat{a}\right) - \frac{1}{2\pi i}\oint_{|z'|<1}\frac{\mathrm{d}z'}{z'-z}\frac{z^{'2}-1}{z^{'2}}\widetilde{\mathcal{M}}\left(z',\hat{a}\right)\,,
\end{equation}
we can obtain the following dispersion relation in the complex plane 
\begin{equation}\label{2ch-z-disp}
	\widetilde{\mathcal{M}}\left(z,\hat{a}\right) = \widetilde{\mathcal{M}}\left(0,\hat{a}\right) + \frac{1}{\pi}\frac{z^{2}}{1-z^{2}}\int_{|z'|=1}\frac{\mathrm{d}z'}{z'}\frac{z^{'2}-1}{z^{'2}-z^{2}}\widetilde{\mathcal{A}}\left(z',\hat{a}\right).
\end{equation}
Written in terms of the kinematic variables, \eqref{2ch-z-disp} becomes
\begin{equation}\label{nonloc-para-disp}
	\mathcal{M}\left(s_{1}, s_{2}\right) = \mathcal{M}\left(-\lambda,-\lambda\right) + \frac{1}{\pi}\int\frac{\mathrm{d}\sigma}{\sigma+\lambda}\left[\frac{s_{1}+\lambda}{\sigma-s_{1}}+\frac{s_{2}+\lambda}{\sigma-s_{2}}\right]\mathcal{A}^{(s_{1})}\left(\sigma,\frac{\hat{a}\left(\sigma+\lambda\right)}{\sigma+\lambda-\hat{a}}-\lambda\right).
\end{equation}
This is the parametric version of the two two-channel symmetric dispersion relation. 

We define
\begin{equation}
	\hat{x} = s_{1}+s_{2}+2\lambda, \qquad \hat{y}= \left(s_{1}+\lambda\right)\left(s_{2}+\lambda\right).
\end{equation}
Now using \eqref{a-mod}, we see that 
\begin{equation}
	\hat{y} = \hat{a}\hat{x}.
\end{equation}
Note that the kernel in \eqref{nonloc-para-disp} can be written as
\begin{eqnarray}
	\frac{\hat{x}\left(\sigma+\lambda\right)-2\hat{y}}{\left(\sigma+\lambda\right)^{2}-\hat{x}\left(\sigma+\lambda\right)+\hat{y}} &=& \frac{\left(\sigma+\lambda\right)^{2}-\hat{y}}{\left(\sigma+\lambda\right)^{2}-\hat{x}\left(\sigma+\lambda\right)+\hat{y}}-1 \nonumber\\
	&= & \frac{\left(\sigma+\lambda\right)^{2}-\hat{y}}{\left(\sigma+\lambda\right)^{2}+\hat{y}}\sum_{k=0}^{\infty}\left(\frac{\hat{x}\left(\sigma+\lambda\right)}{\left(\sigma+\lambda\right)^{2}+\hat{y}}\right)^{k}.
\end{eqnarray}
When a monomial in $\hat{a}$ acts on this kernel, we will keep the local terms by removing the negative powers of $\hat{x}$ generated in the process. This leads to effectively changing 
\begin{equation}
	\hat{a}\rightarrow \frac{\hat{y}\left(\sigma+\lambda\right)}{\left(\sigma+\lambda\right)^{2}+\hat{y}} .
\end{equation}
This implies that $s_{2}$ becomes
\begin{equation}
	s_{2}\left(\sigma\right)=\frac{\hat{y}}{\sigma+\lambda}-\lambda .
\end{equation}
Therefore, finally, we obtain the local dispersion relation
\begin{eqnarray}
	\mathcal{M}\left(s_{1},s_{2}\right) 
	&=& \mathcal{M}\left(-\lambda,-\lambda\right)-\frac{1}{\pi}\int\mathrm{d}\sigma\left[\frac{1}{s_{1}-\sigma}+\frac{1}{s_{2}-\sigma}+\frac{1}{\sigma+\lambda}\right]\mathcal{A}\left(\sigma,\frac{\left(s_{1}+\lambda\right)\left(s_{2}+\lambda\right)}{\sigma+\lambda}-\lambda\right) \nonumber\\
	&& \hspace{2.5cm}- \frac{1}{\pi}\int\frac{\mathrm{d}\sigma}{\sigma+\lambda}\mathcal{A}\left(\sigma,-\lambda\right) .
\end{eqnarray}

\section{Subtracted dispersion relation}
\label{app:HS}
The higher subtracted dispersion relation given in \eqref{HS-method1} can be obtained by considering the following contour integral,
\begin{eqnarray}\label{HS-cont}
    &&\int_{\Gamma}\frac{\mathrm{d}\sigma}{2\pi i} \left[ \mathcal{H}^{(\lambda)}\left(\sigma,s,t\right)\mathcal{M}\left(\sigma,\frac{\left(s+\lambda\right)\left(t+\lambda\right)}{\lambda+\sigma}-\lambda\right) \right.
    \nonumber\\
    && \left. - \sum_{p,q}\frac{\partial^{p}}{\partial x^{p}}\frac{\partial^{q}}{\partial y^{q}}\Biggl\{\left(\frac{1}{\lambda +\sigma }+\frac{x-2 \sigma }{\sigma ^2-\sigma  x+y}\right)\mathcal{\mathcal{M}}\left(\sigma,\frac{y + \lambda x+\lambda^{2}}{\sigma+\lambda}-\lambda\right)\Biggr\}\right]
\end{eqnarray}
   Here $x=s+t$ and $y=st$. The sum over $p$ and $q$ should be chosen such that the integrand falls off faster than $\frac{1}{\sigma}$ at large $\sigma$. From the first line, we get the usual contributions,
    \begin{eqnarray}\label{highersub-1}
        && -2\mathcal{M}\left(s,t\right) +\mathcal{M}\left(-\lambda,\infty\right) \nonumber\\
        && -\frac{1}{\pi}\int_{\mathcal{C}_{1}}\mathrm{d}\sigma\left(\frac{1}{s-\sigma}+\frac{1}{t-\sigma}+\frac{1}{\sigma+\lambda}\right)\mathcal{A}^{(s)}\left(\sigma,\frac{\left(s+\lambda\right)\left(t+\lambda\right)}{\sigma+\lambda}-\lambda\right)\nonumber\\
        && -\frac{1}{\pi}\int_{\mathcal{C}_{2}}\mathrm{d}\sigma\left(\frac{1}{s-\sigma}+\frac{1}{t-\sigma}+\frac{1}{\sigma+\lambda}\right)\mathcal{A}^{(t)}\left(\sigma,\frac{\left(s+\lambda\right)\left(t+\lambda\right)}{\sigma+\lambda}-\lambda\right).
    \end{eqnarray}
Here we have assumed, without justification, the presence of a simple at $\sigma=-\lambda$. In actual cases, $\sigma=-\lambda$ may correspond to either a higher-order pole or a branch point. In general, the amplitude can grow at large $s$ or $t$ and therefore $\mathcal{M}\left(-\lambda,\infty\right)$ diverges. Later, we will see that this term gets cancelled when we include contributions from the second line of \eqref{HS-cont}. 

There are two types of singularities in the second line:
\begin{itemize}
    \item Kernel part has simple poles at  $\sigma^{\pm}=\frac{1}{2}\left(x\pm\sqrt{x^{2}-4y}\right)$. Note that $\sigma^{\pm}$ are the solutions of $s_{1}$ and $s_{2}$ which satisfy $s_{1}+s_{2}=x$ and $s_{1}s_{2}=y$. In addition to these poles, we assume that $\sigma=-\lambda$ is a simple pole. 
    \item Amplitude has singularities in $\sigma$ in the $s$ and $t$ channels. 
\end{itemize}
Contribution from the kernel part is proportional to
\begin{eqnarray}\label{highersub-2}
    &&\frac{\partial^{p}}{\partial x^{p}}\frac{\partial^{q}}{\partial y^{q}}\Biggl\{\left[\mathcal{M}\left(\sigma,\frac{y + \lambda x+\lambda^{2}}{\sigma+\lambda}-\lambda\right)\right]_{\sigma=-\lambda}-\left[\mathcal{M}\left(\sigma,\frac{y + \lambda x+\lambda^{2}}{\sigma+\lambda}-\lambda\right)\right]_{\sigma=\sigma^{+}} \nonumber\\
    && \hspace{2cm}-\left[\mathcal{M}\left(\sigma,\frac{y + \lambda x+\lambda^{2}}{\sigma+\lambda}-\lambda\right)\right]_{\sigma=\sigma^{+}}\Biggr\}_{x=-2\lambda,\; y=\lambda^{2}}\nonumber\\
    & = & \frac{\partial^{p}}{\partial x^{p}}\frac{\partial^{q}}{\partial y^{q}}\Biggl\{\mathcal{M}\left(-\lambda,\infty\right)-2\mathcal{M}\left(\sigma^{+},\sigma^{-}\right)\Biggr\}_{x=-2\lambda,\; y=\lambda^{2}} \nonumber\\
    & = & \mathcal{M}\left(-\lambda,\infty\right)\delta_{p,0}\delta_{q,0} - 2W_{pq}\left(-2\lambda,\lambda^{2}\right).
\end{eqnarray}
 Any derivative acting on $\mathcal{M}\left(-\lambda,\infty\right)$ vanishes.
From the singularities of the amplitude, we will have 
\begin{eqnarray}\label{highersub-3}
    &&\frac{\partial^{p}}{\partial x^{p}}\frac{\partial^{q}}{\partial y^{q}}\Biggl\{\int_{\mathcal{C}_{1}}\mathrm{d}\sigma\left(\frac{1}{\lambda +\sigma }+\frac{x-2 \sigma }{\sigma ^2-\sigma  x+y}\right)\mathcal{A}^{(s)}\left(\sigma,\frac{y + \lambda x+\lambda^{2}}{\sigma+\lambda}-\lambda\right) \nonumber\\
    && + \int_{\mathcal{C}_{1}}\mathrm{d}\sigma\left(\frac{1}{\lambda +\sigma }+\frac{x-2 \sigma }{\sigma ^2-\sigma  x+y}\right)\mathcal{A}^{(t)}\left(\sigma,\frac{y + \lambda x+\lambda^{2}}{\sigma+\lambda}-\lambda\right)\Biggr\}_{x=-2\lambda,\; y=\lambda^{2}}
\end{eqnarray}
Crossing symmetry implies that the two integrals are equal. Collecting all the terms with their relative signs taken into account, we obtain \eqref{HS-method1}.

\section{Details of numerics and SDPB implementation}
As discussed in subsection \ref{g2g3bounds}, to obtain a lower bound on $W_{0,1}$ given any $W_{1,0}$ in the allowed range, we need to solve the following optimisation problem.
\begin{empheq}[box=\boxed]{align}
    &\text{Given $W_{1,0}$, find $\{\alpha(\lambda_i),\beta(\lambda_i)\}$ that maximize} \sum_{\lambda_i \in \lambda_{grid}} 
        \alpha(\lambda_i)\frac{1}{\lambda_i^2}
        - \beta(\lambda_i)\left(W_{1,0} - \frac{2}{\lambda_i}\right), \nonumber\\[4pt]
    &\text{with } 
      \sum_{\lambda_i \in \lambda_{grid}} \alpha(\lambda_i) = 1 
      \text{ and } 
      \sum_{\lambda_i \in \lambda_{grid}} 
        \alpha(\lambda_i)\,\tilde{w}_{0,1}(\ell,k,\lambda_i) 
        + \beta(\lambda_i)\,\tilde{w}_{1,0}(\ell,k,\lambda_i)\ge 0, \nonumber\\[4pt]
    &\nonumber \text{for } 1 \le k \le k_{\text{max}}, \quad 0 \le \ell \le 2k+2.
\end{empheq}
Similarly, we solve an analogous optimisation problem to obtain an upper bound on $W_{0,1}$. Since the bounds are straight lines, we only need to solve the optimisation problem for two choices of $W_{1,0}$. \\
We choose 
\begin{equation}
\lambda_{grid} = 1.01 +\frac{n}{2}, \quad   n \in \{ 0,1,2,\cdots, n_{max}\}\,.
\end{equation}
To obtain the bounds, $n_{max}$ must be increased as $k_{max}$ is increased. For Fig. \ref{fig:g2g3plot}, we required the following $(k_{max}, n_{max})$ pairs:
\begin{equation}
  (k_{max}, n_{max}) \in   \{(50,25), ~(100,48),~ (150,71),~ (200,93),~(250,116) \}\,.
\end{equation}
We imposed the constraints using SDPB 3.0.0 \cite{SDPB}. For all the runs, we chose the following SDPB parameters: "––precision=1024" and "––dualityGapThreshold=1e-10".

\end{document}